\documentclass[twocolumn,english,prb,showpacs,preprintnumbers,amsmath,amssymb]{revtex4}
\usepackage[latin9]{inputenc}
\usepackage{graphicx}
\usepackage{dcolumn}
\usepackage{bm}
\usepackage{color}

\begin{document}

\title{Effect of C-face $4H$-SiC(0001) deposition on thermopower of single
and multilayer graphene in AA, AB and ABC stacking}

\author{Ma\l{}gorzata~Wierzbowska}
\email{malgorzata.wierzbowska@fuw.edu.pl}
\affiliation{Institute of Theoretical Physics, Faculty of Physics, University
of Warsaw, ul. Ho\.{z}a 69, 00-681 Warszawa, Poland }

\author{Adam Dominiak}
\email{adam.dominiak@itc.pw.edu.pl}
\affiliation{Institute of Heat Engineering, Faculty of Power and Aeronautical
Engineering, Warsaw University of Technology, ul. Nowowiejska 21/25,
00-665 Warszawa, Poland}

\author{Giovanni Pizzi}
\affiliation{Theory and Simulation of Materials, \'Ecole Polytechnique F\'ed\'erale
de Lausanne, 1015 Lausanne, Switzerland}

\begin{abstract}
The Seebeck coefficient in multilayer graphene is investigated within
the density-functional theory, using the semiclassical Boltzmann equations and
interpolating the bands in a maximally-localized Wannier functions basis set. 
We compare various graphene stackings (AA, AB and ABC) both free-standing
and deposited on a $4H$-SiC(0001) C-terminated
substrate. We find that the presence of the SiC substrate
can significantly affect the thermopower properties of graphene layers, depending on
the stacking, providing a promising way to
tailor efficient graphene-based devices. %
\end{abstract}



\maketitle

\begin{figure*}
\leftline{ \includegraphics[scale=0.36]{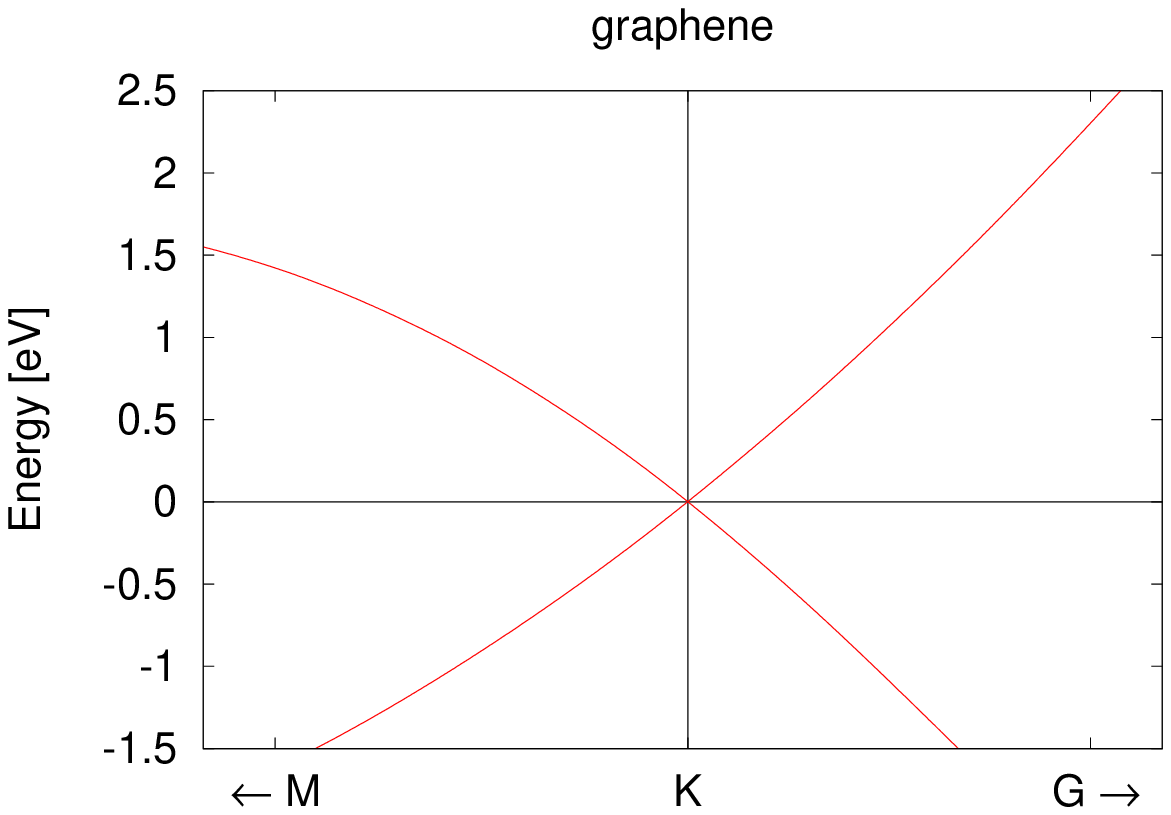} 
\hspace{2mm} \includegraphics[scale=0.11]{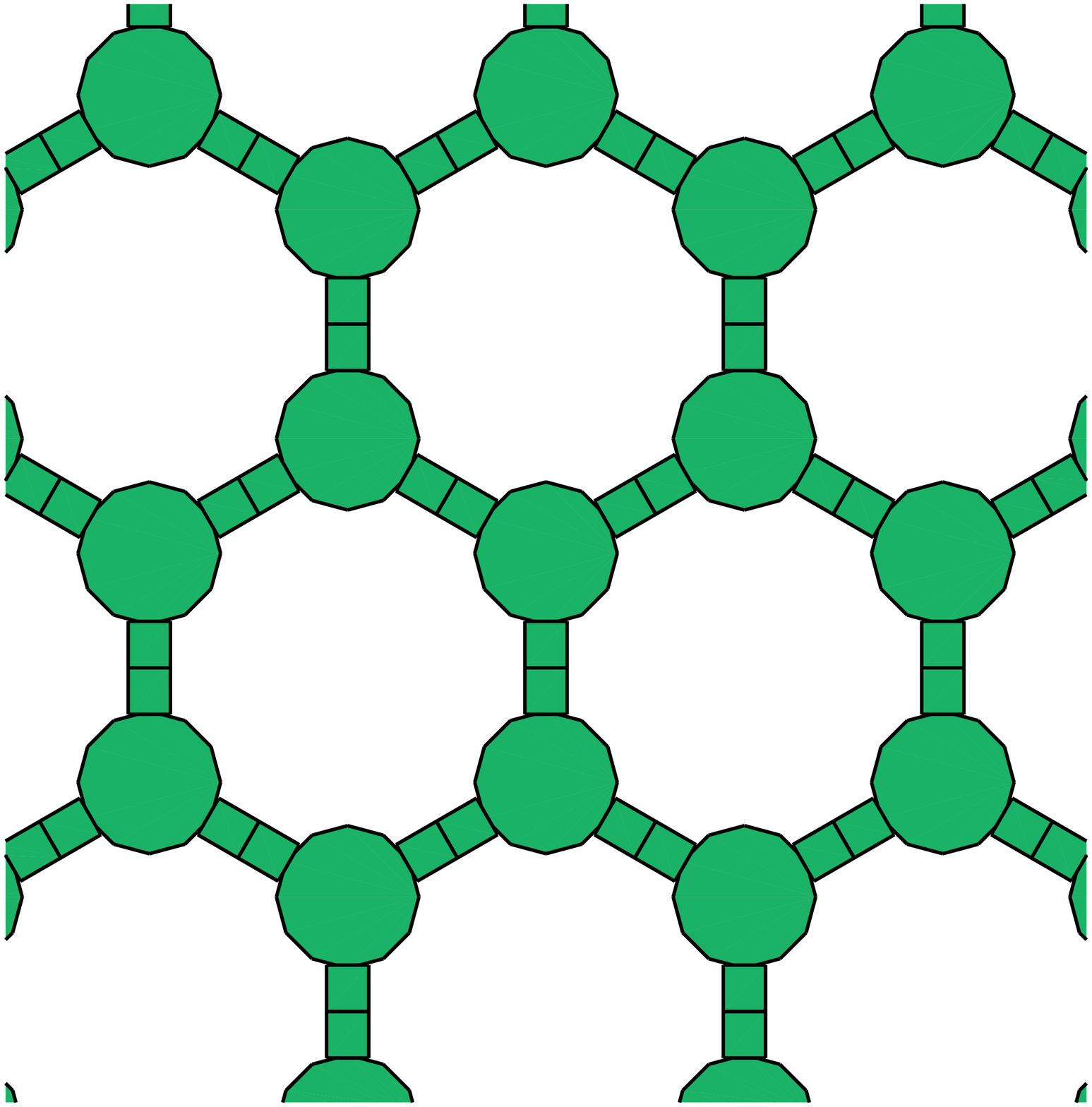}
\hspace{2mm} \includegraphics[scale=0.36]{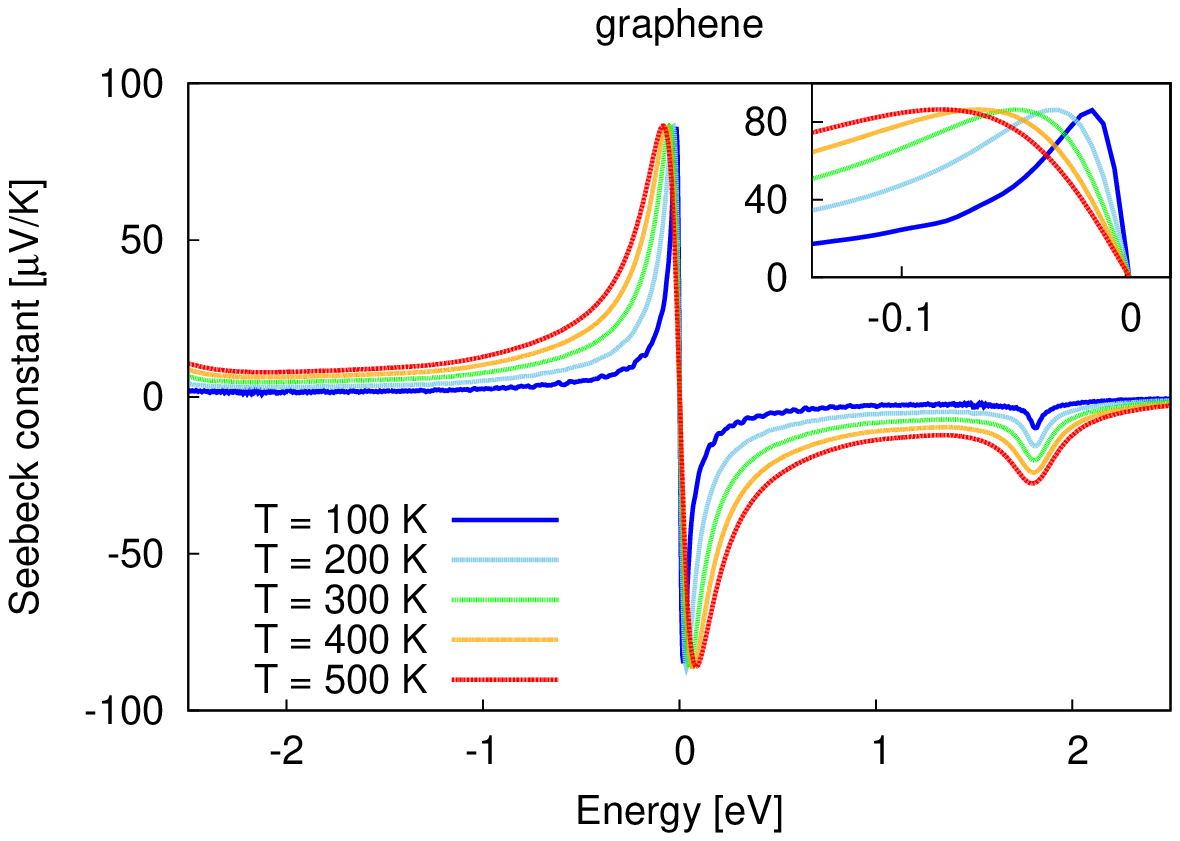}} \vspace{2mm}
 \leftline{ \includegraphics[scale=0.36]{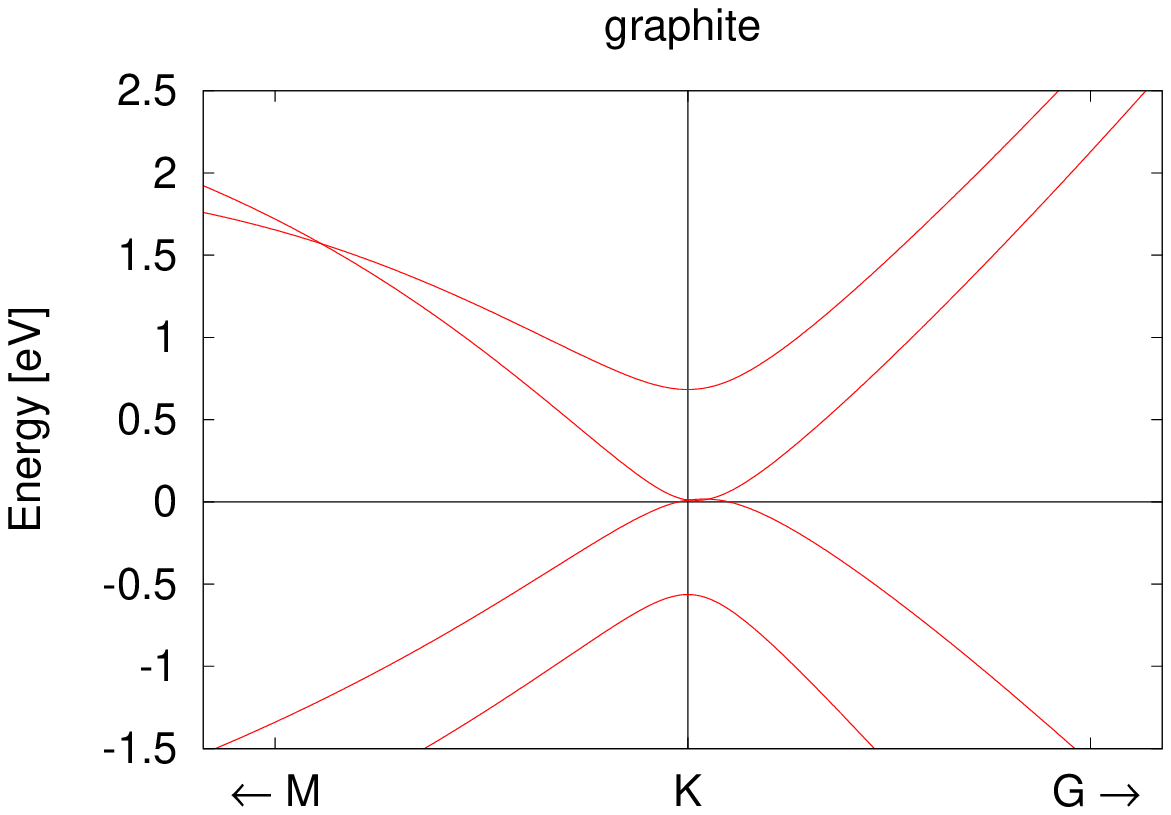} \hspace{2mm}
\includegraphics[scale=0.11]{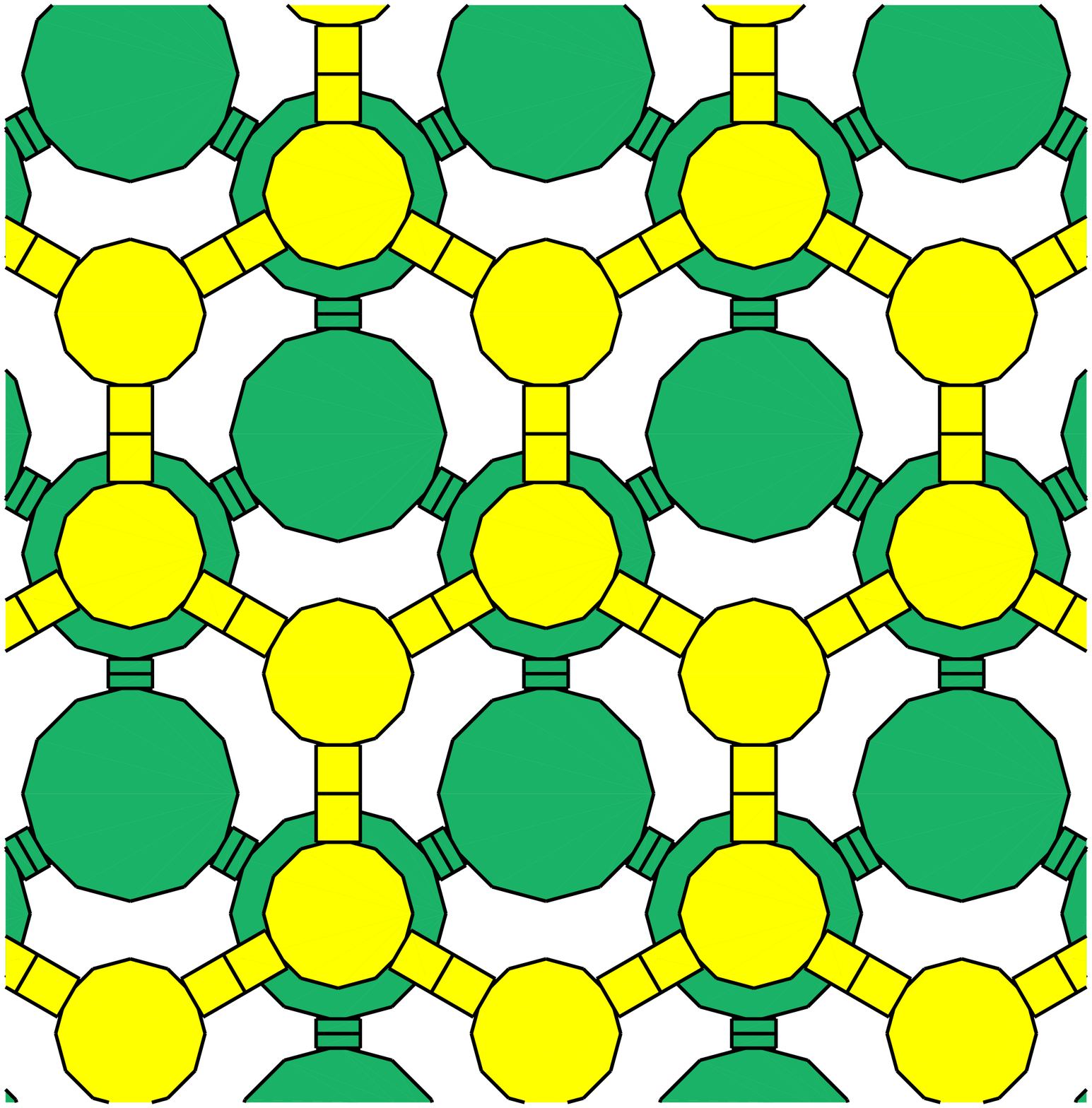} \hspace{2mm} 
\includegraphics[scale=0.36]{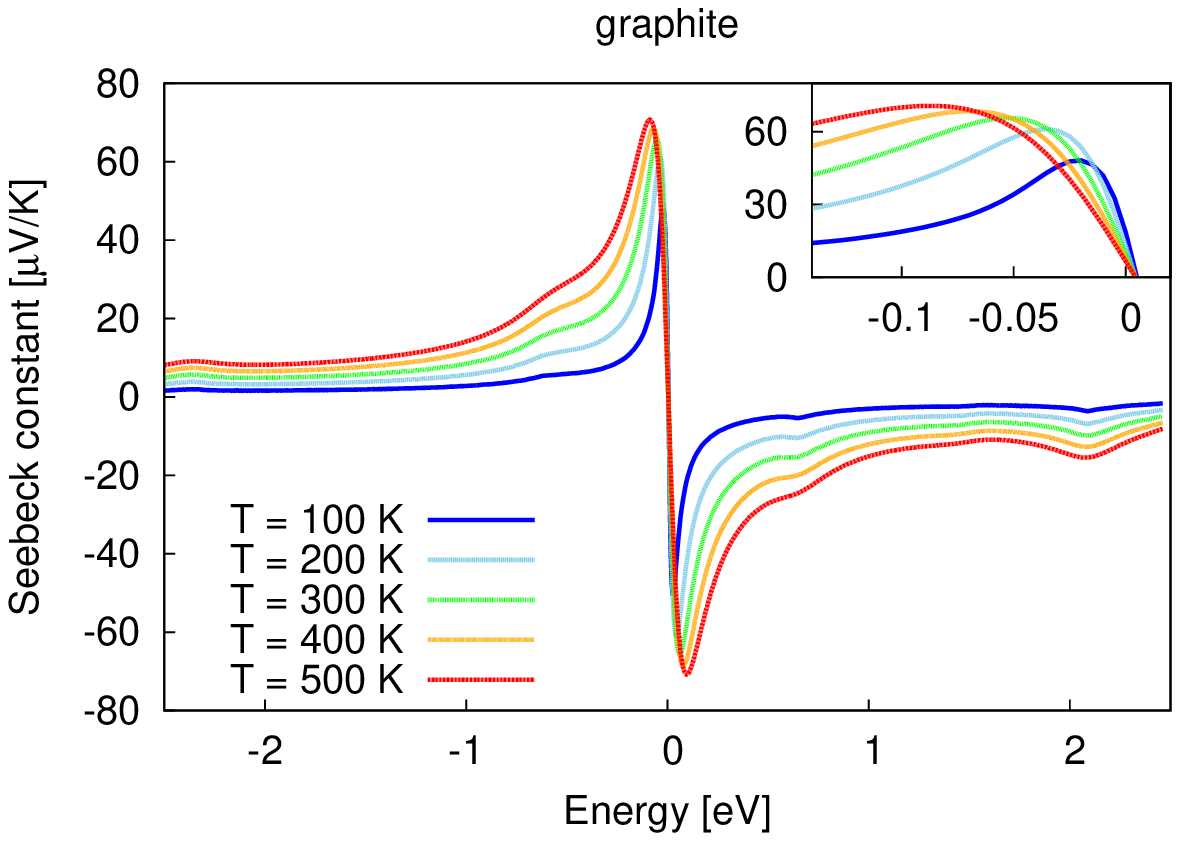}}
\vspace{2mm}
 \leftline{\includegraphics[scale=0.36]{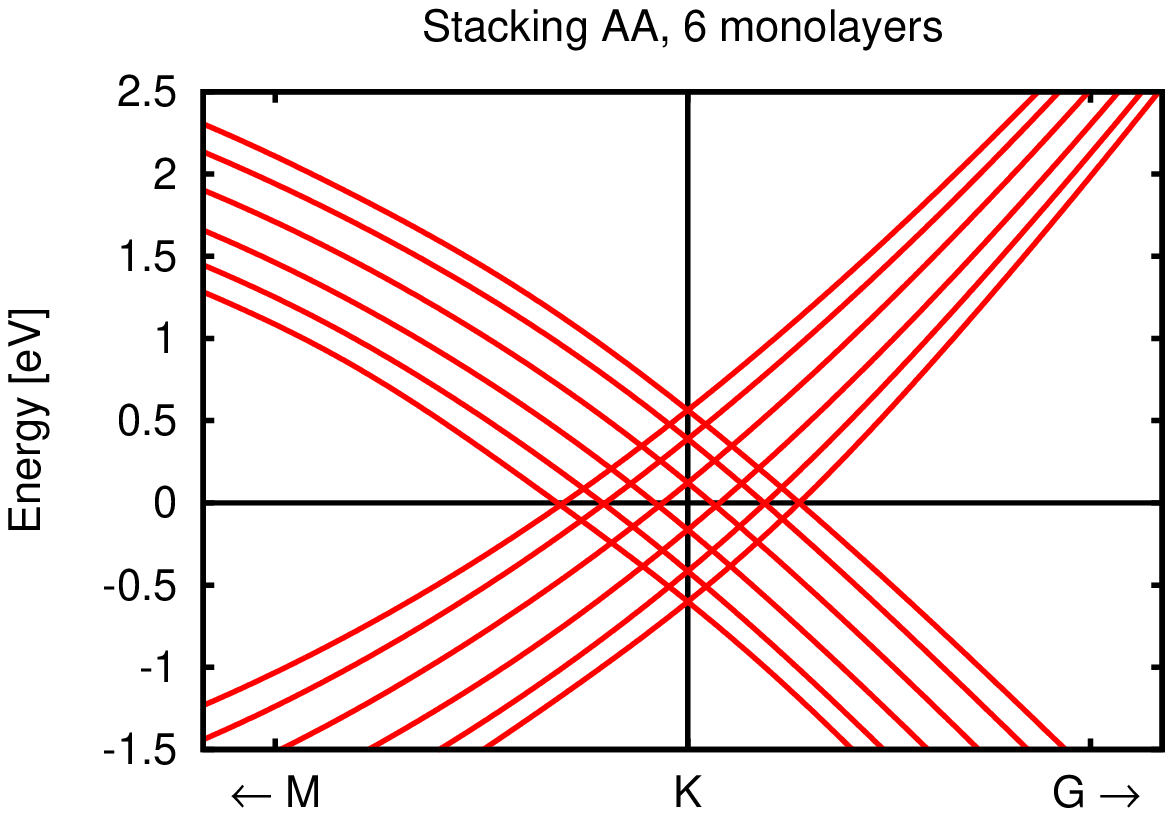} \hspace{2mm} 
\includegraphics[scale=0.11]{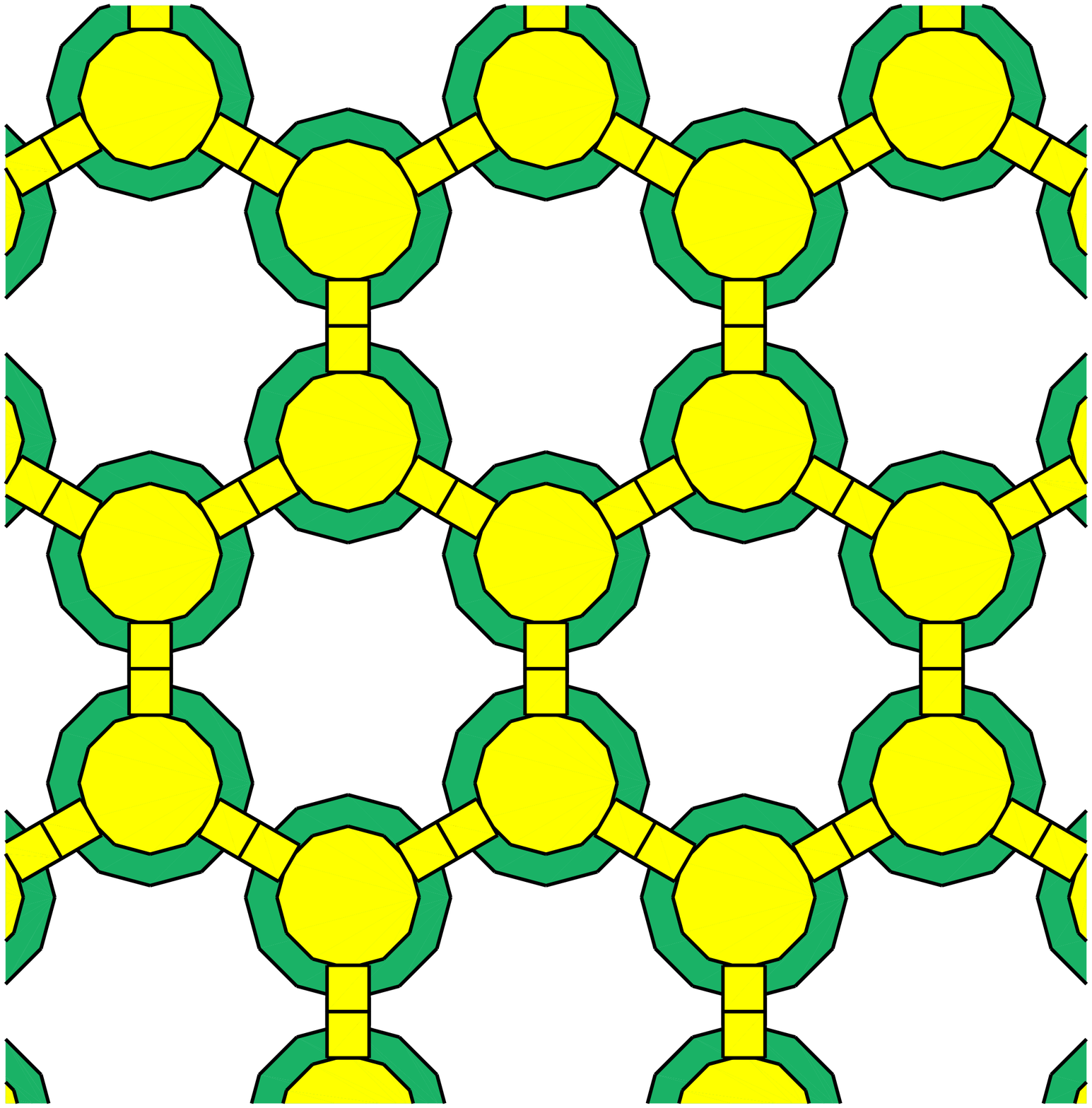}
\hspace{2mm} \includegraphics[scale=0.36]{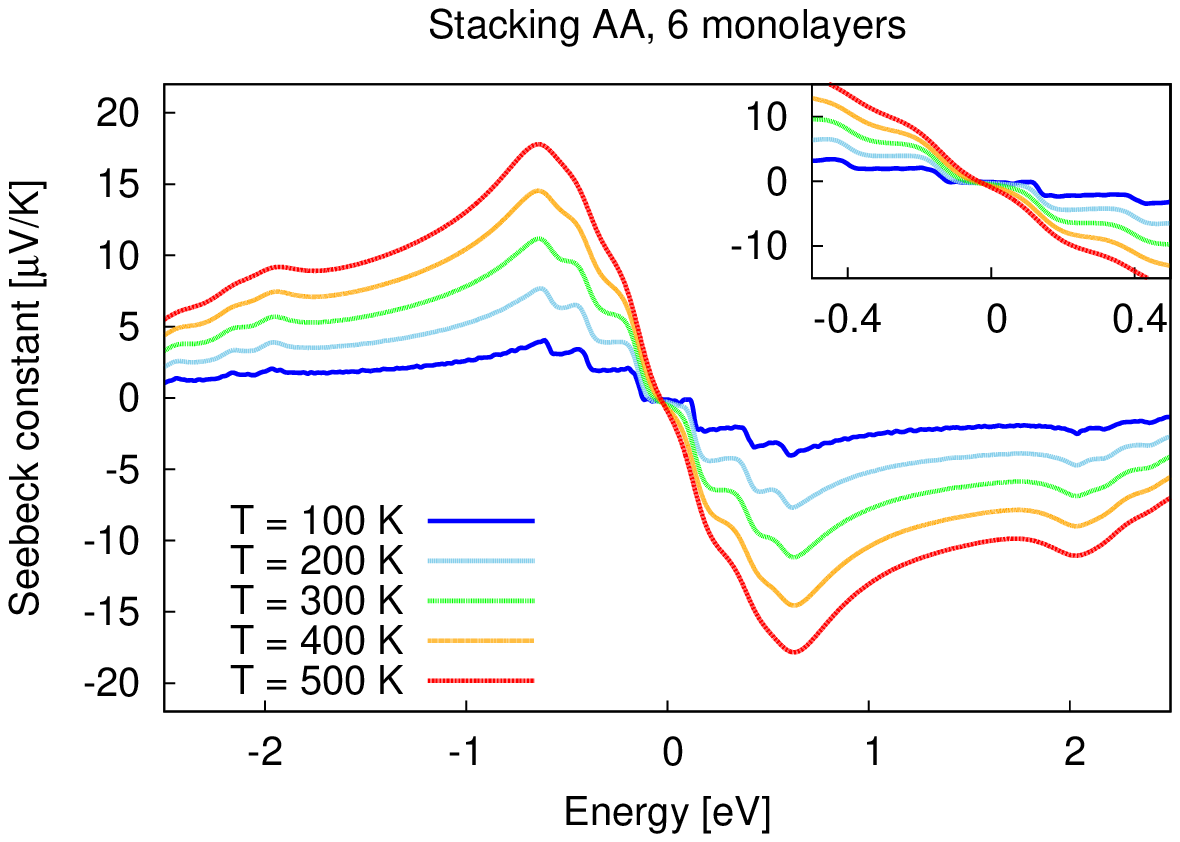} \hspace{2mm} 
\includegraphics[scale=0.36]{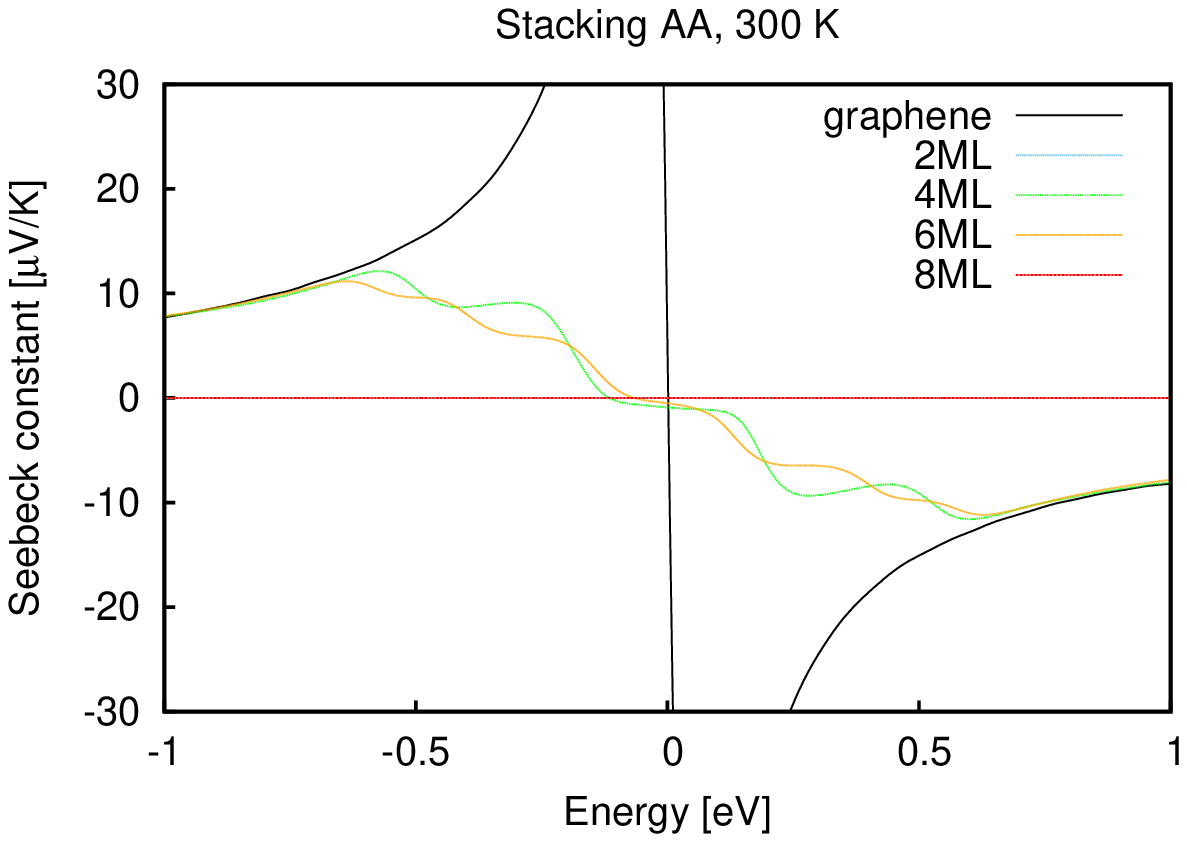}}
\vspace{2mm}
 \leftline{ \includegraphics[scale=0.36]{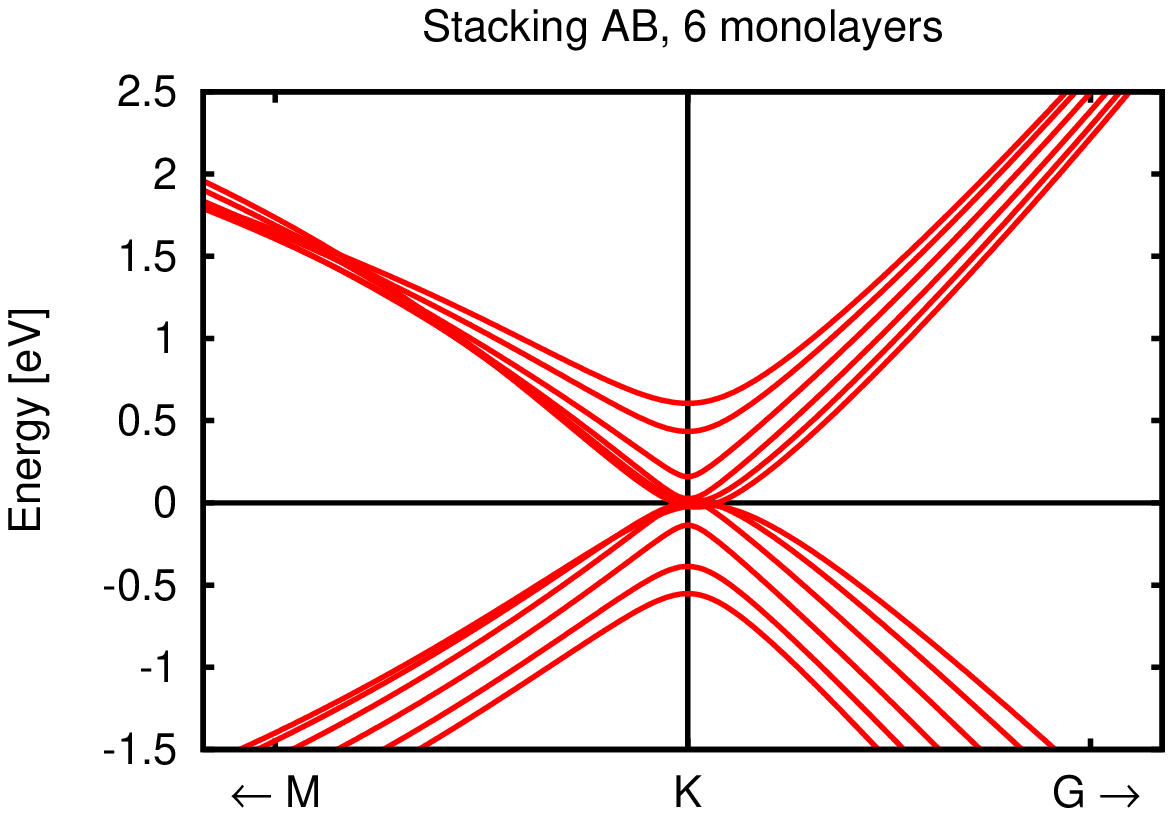} \hspace{2mm} 
\includegraphics[scale=0.11]{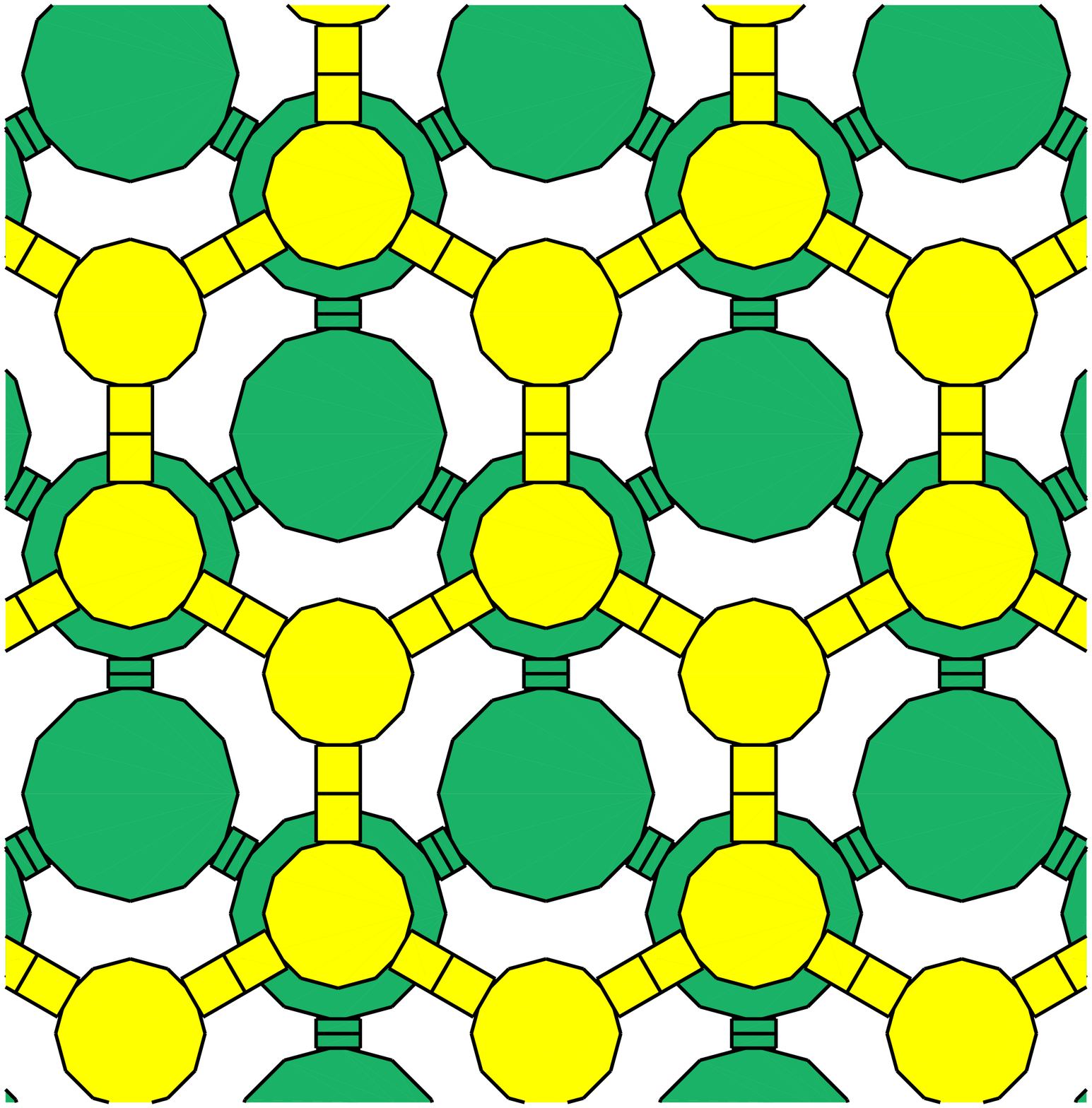}
\hspace{2mm} \includegraphics[scale=0.36]{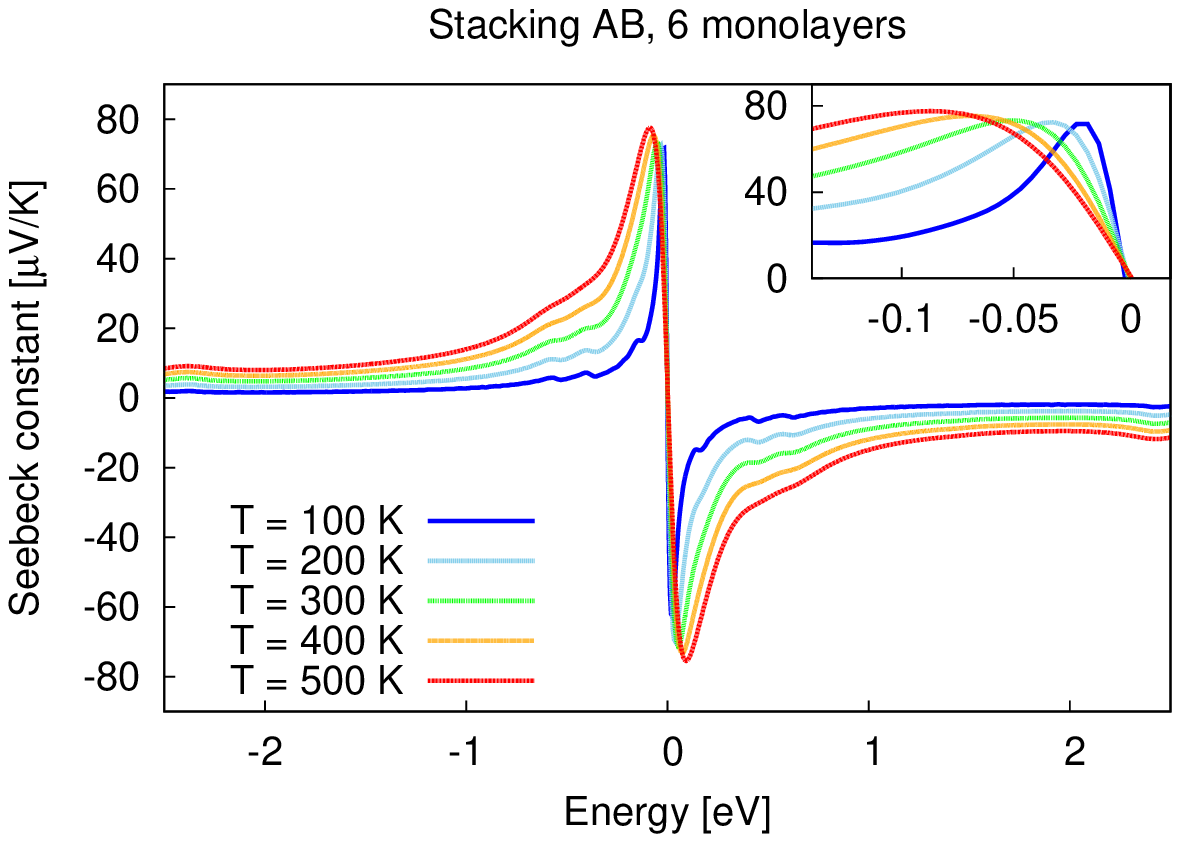} \hspace{2mm} 
\includegraphics[scale=0.36]{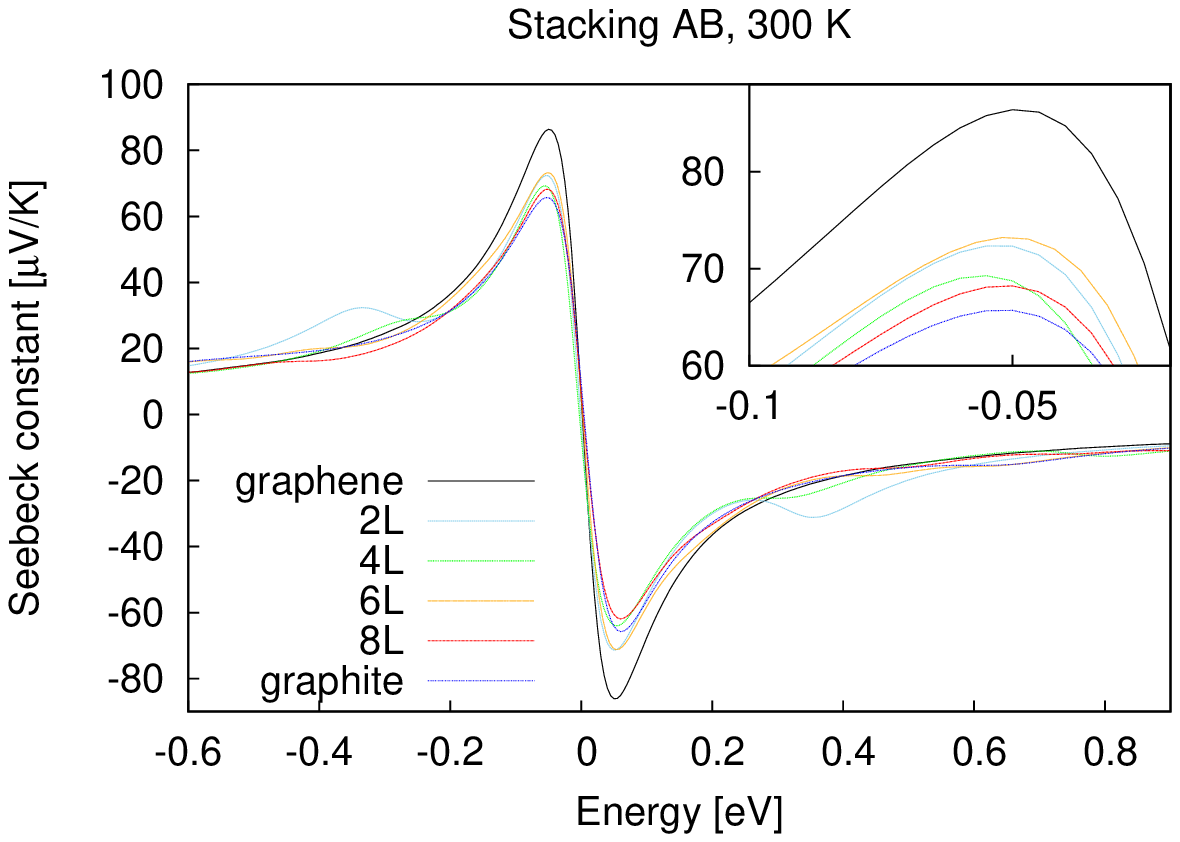}}
\vspace{2mm}
 \leftline{ \includegraphics[scale=0.36]{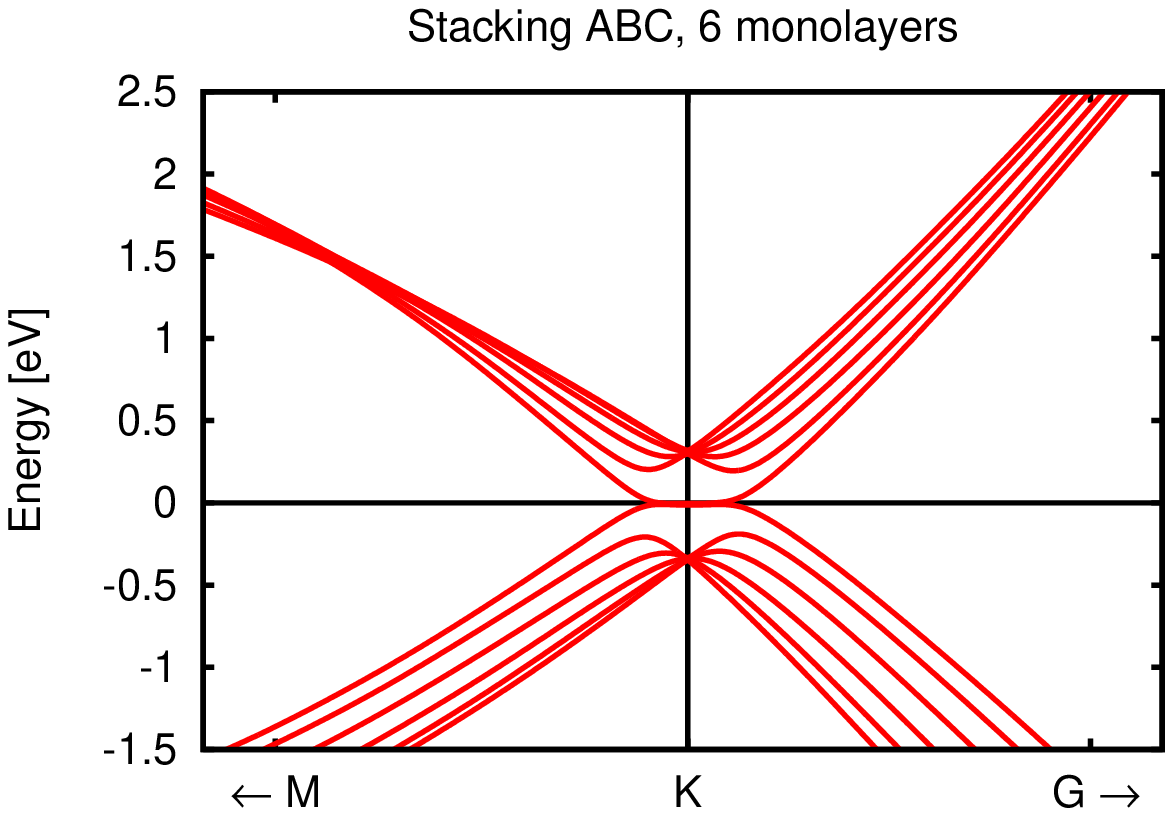} \hspace{2mm} 
\includegraphics[scale=0.11]{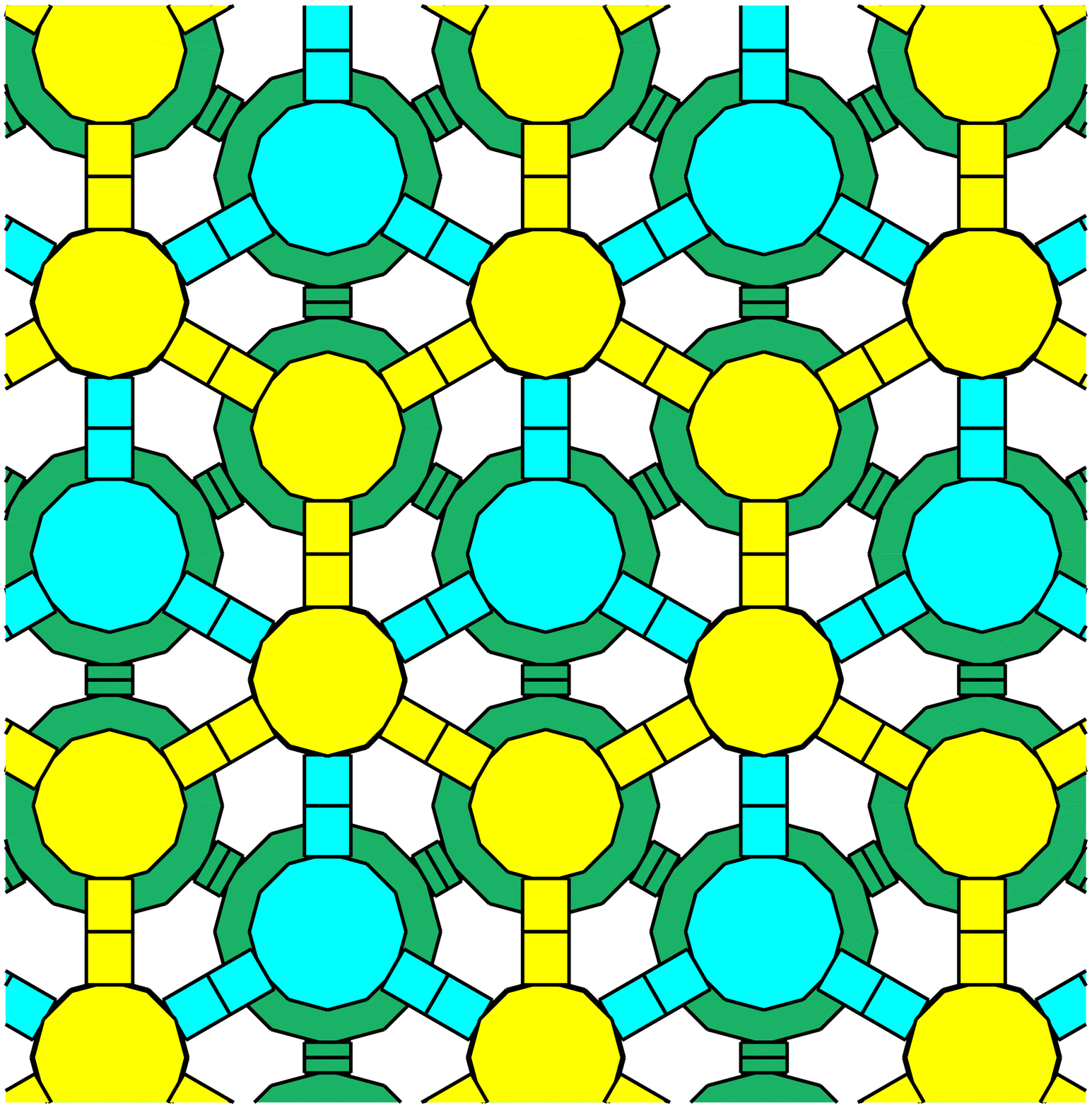}
\hspace{2mm} \includegraphics[scale=0.36]{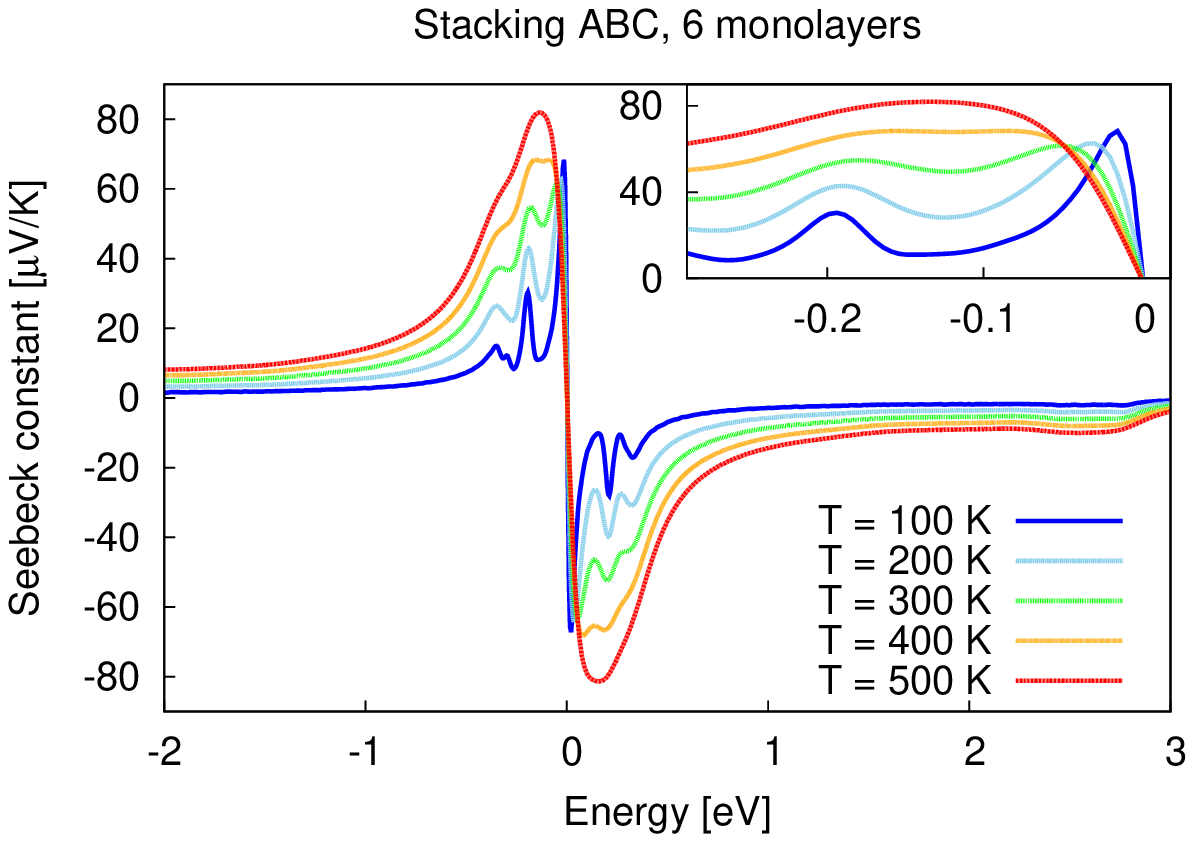} \hspace{2mm} 
\includegraphics[scale=0.36]{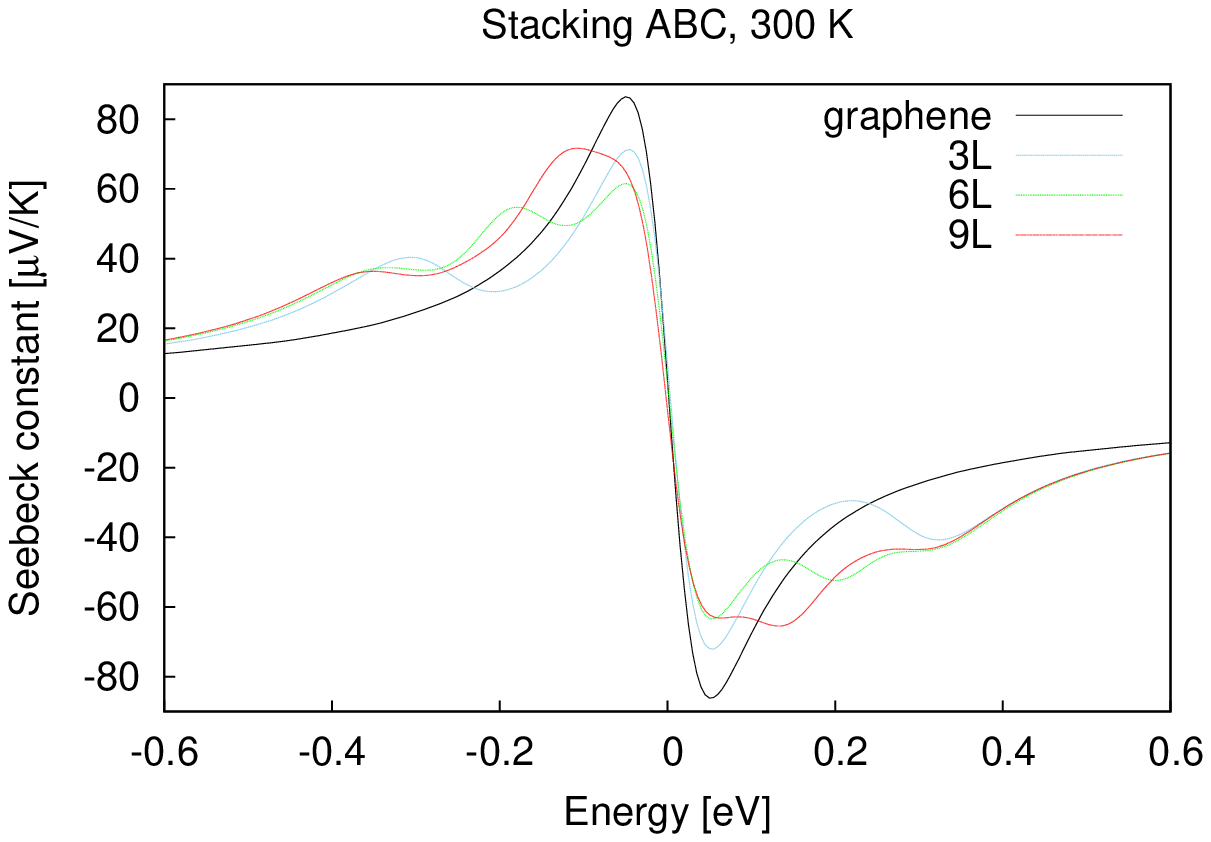}}
\caption{(From left to right) Band structure, top-view geometry, and Seebeck
coefficient (two columns) as a function of chemical potential for
monolayer graphene (first row), graphite (second row), and the AA,
AB and ABC stacked graphene (third, fourth and fifth rows, respectively).
In the third column, the Seebeck coefficient is shown as a function
of the chemical potential for different temperatures; in the fourth
column, it is shown as a function of the chemical potential for different
monolayer (ML) thicknesses.}
\label{b1} 
\end{figure*}
\begin{figure}
\leftline{ \includegraphics[scale=0.35]{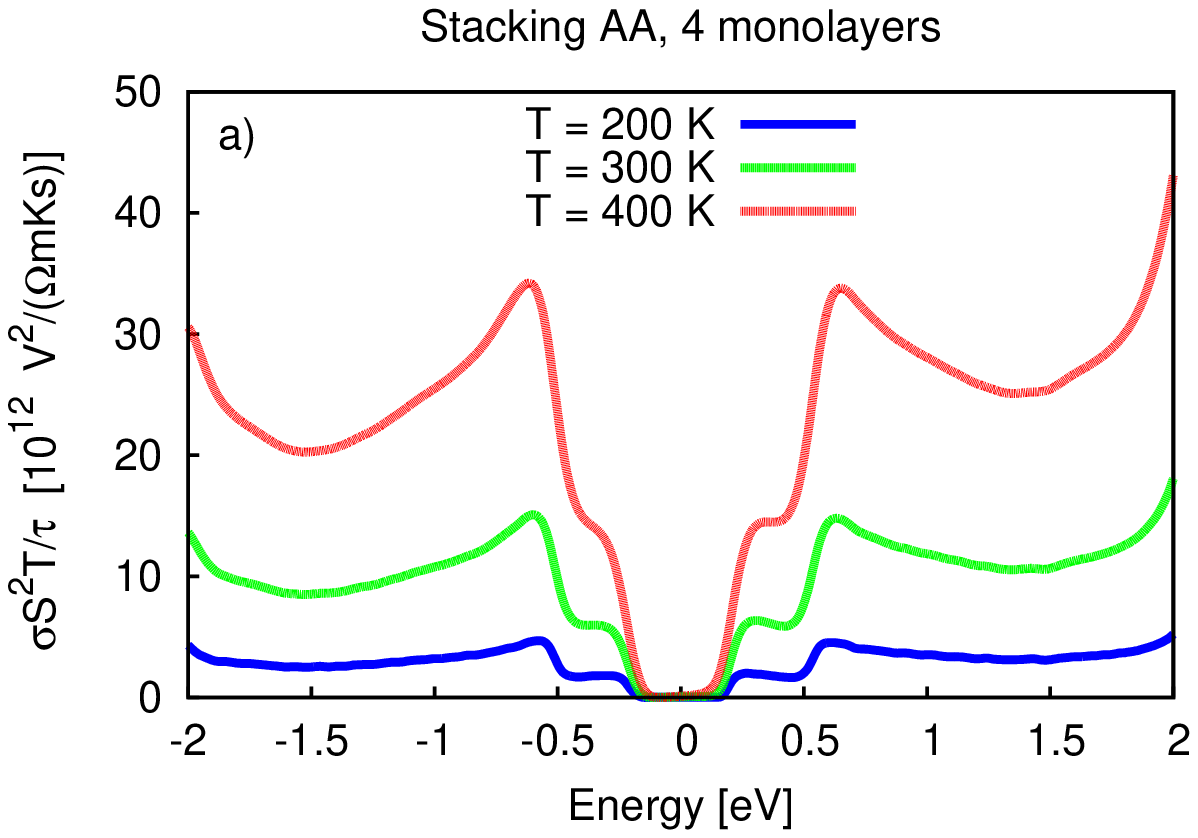} \includegraphics[scale=0.35]{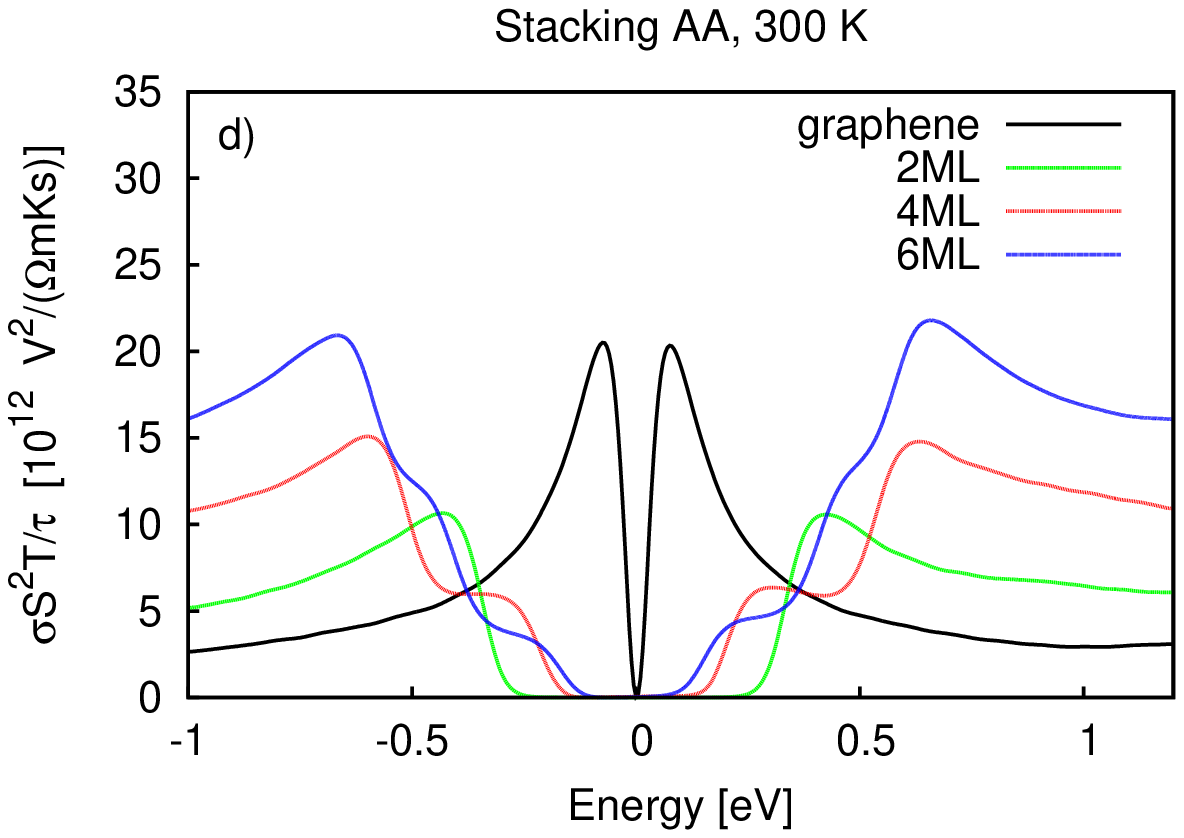}}
\leftline{ \includegraphics[scale=0.35]{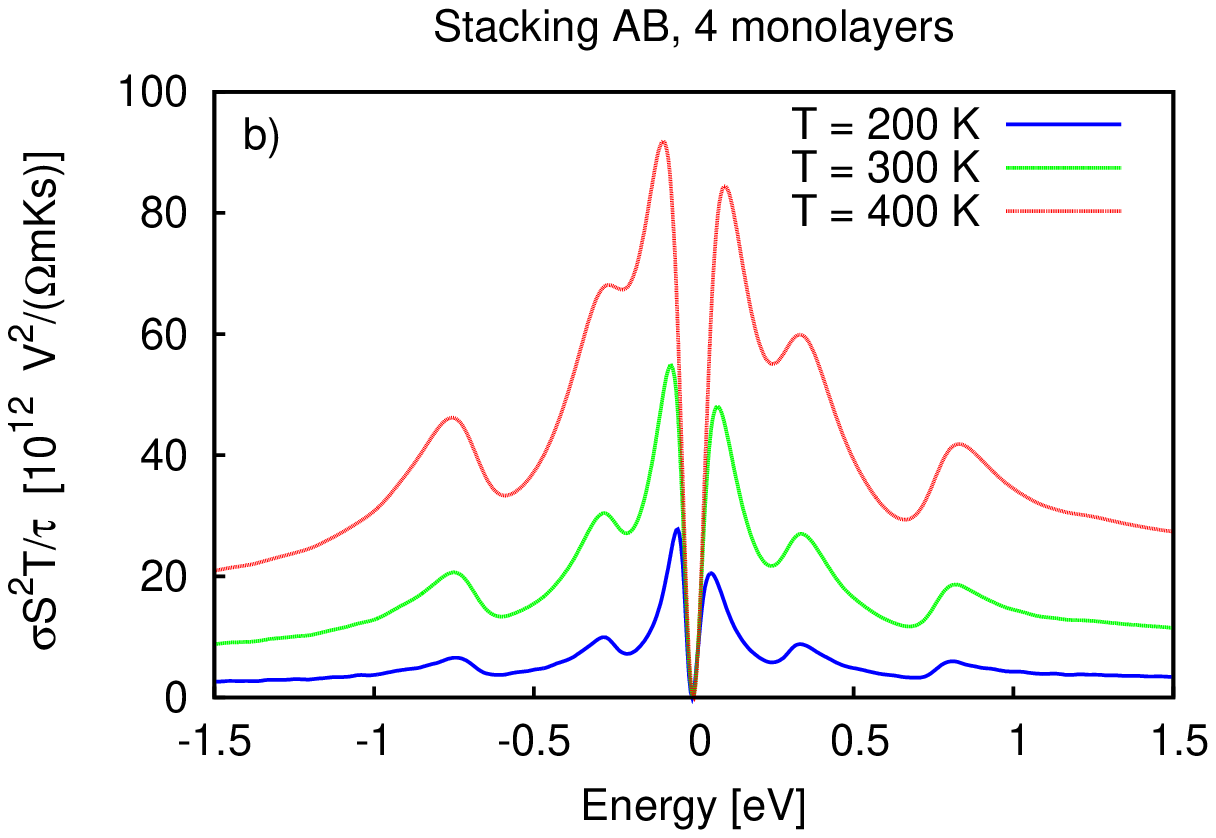} \includegraphics[scale=0.35]{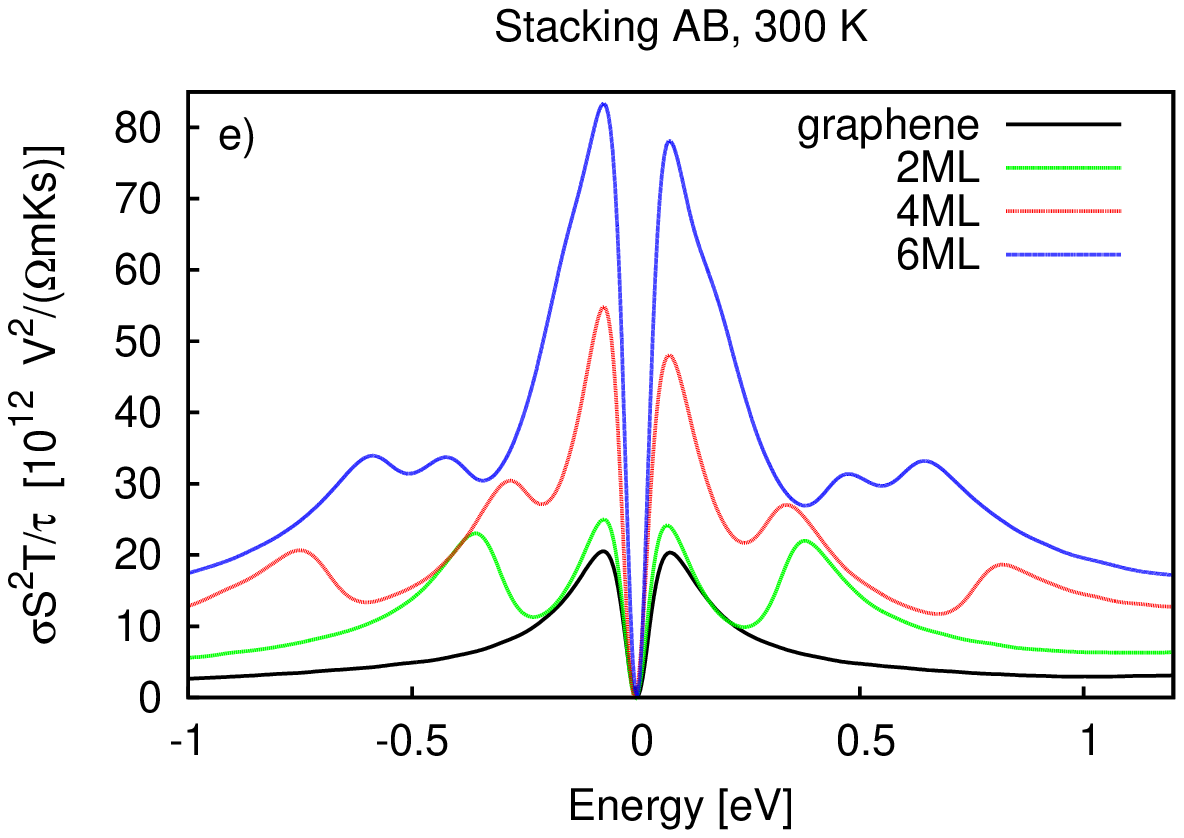}}
\leftline{\includegraphics[scale=0.35]{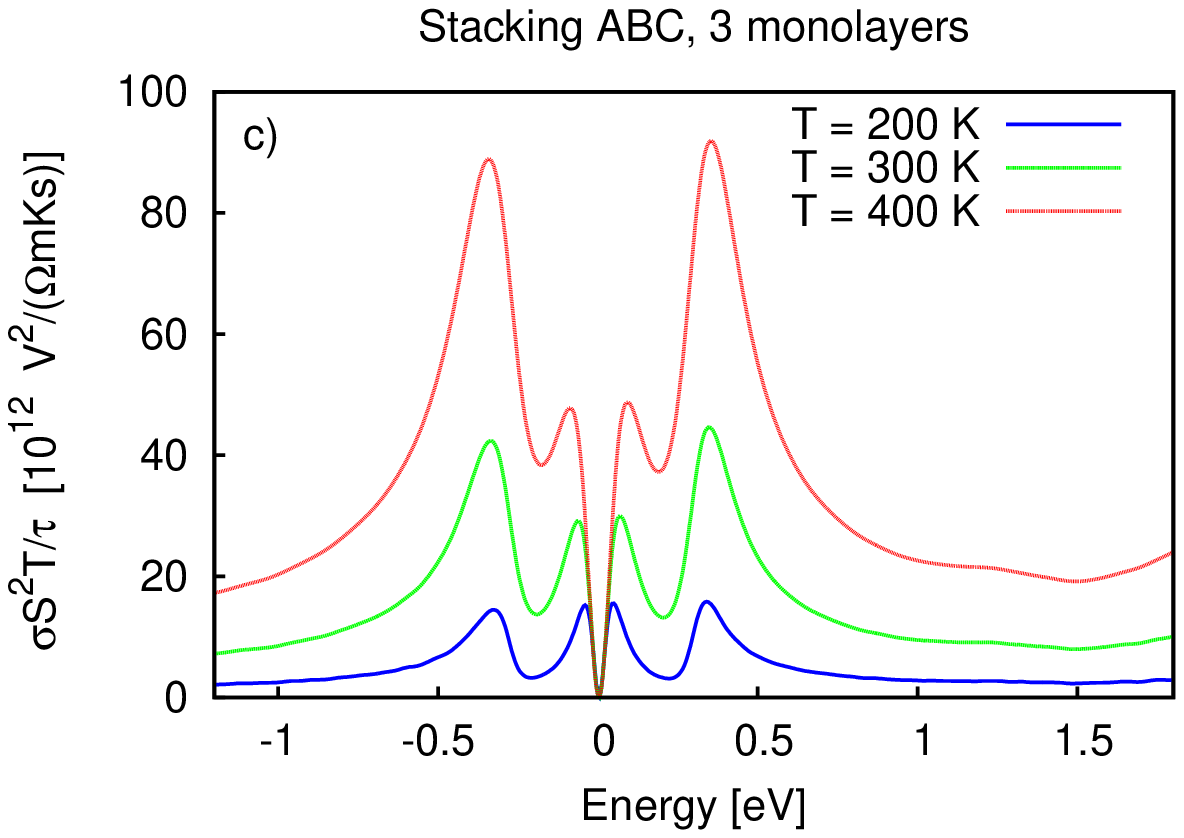} \includegraphics[scale=0.35]{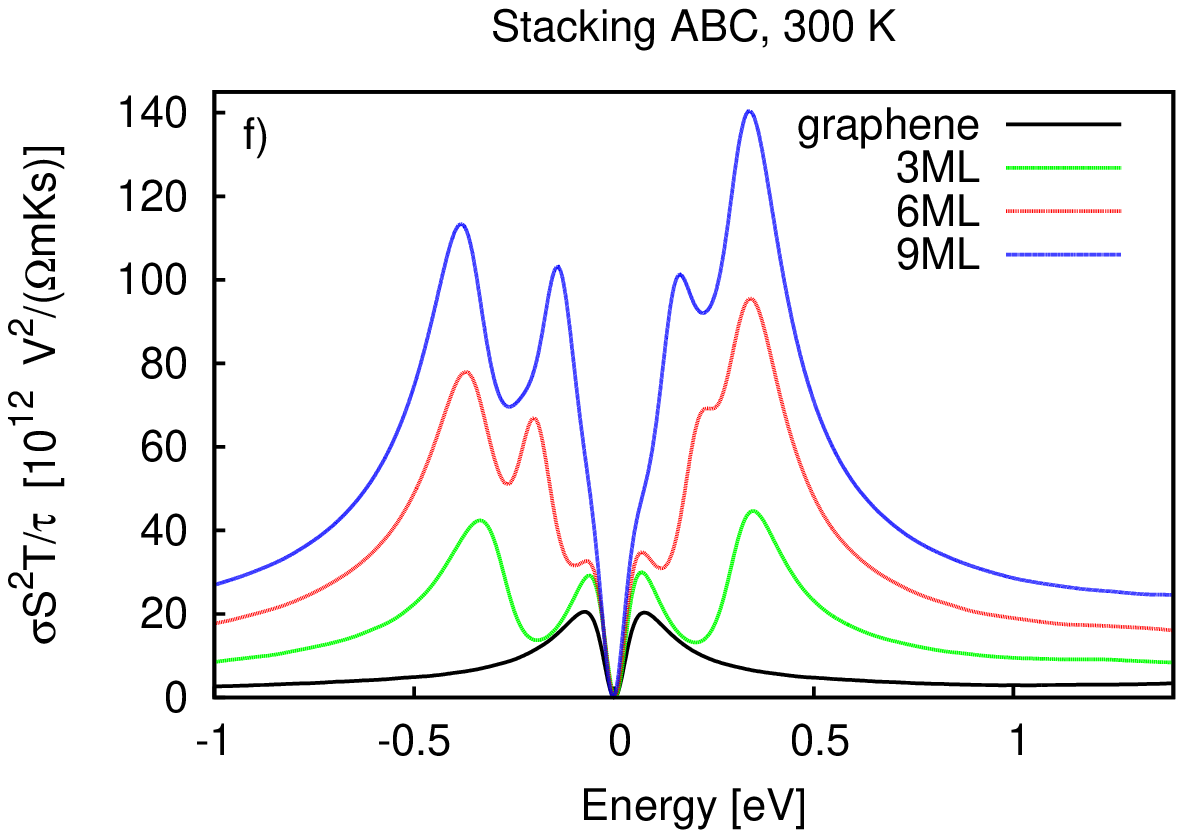}
} \caption{$\sigma S^2 T/\tau$ for different number of free-standing
graphene monolayers (ML)
and types of stacking as a function of the chemical potential, for different
temperatures. 
(a) Four monolayers (4ML) with AA stacking. (b) Four
monolayers with AB stacking. (c) Three monolayers with ABC stacking.
(d) AA, (e) AB and (f) ABC stacking at $T=300K$ for different number
of monolayers. The Fermi energy is set to $E=0$.}
\label{b2} 
\end{figure}
\begin{figure*}
\leftline{\includegraphics[scale=0.45]{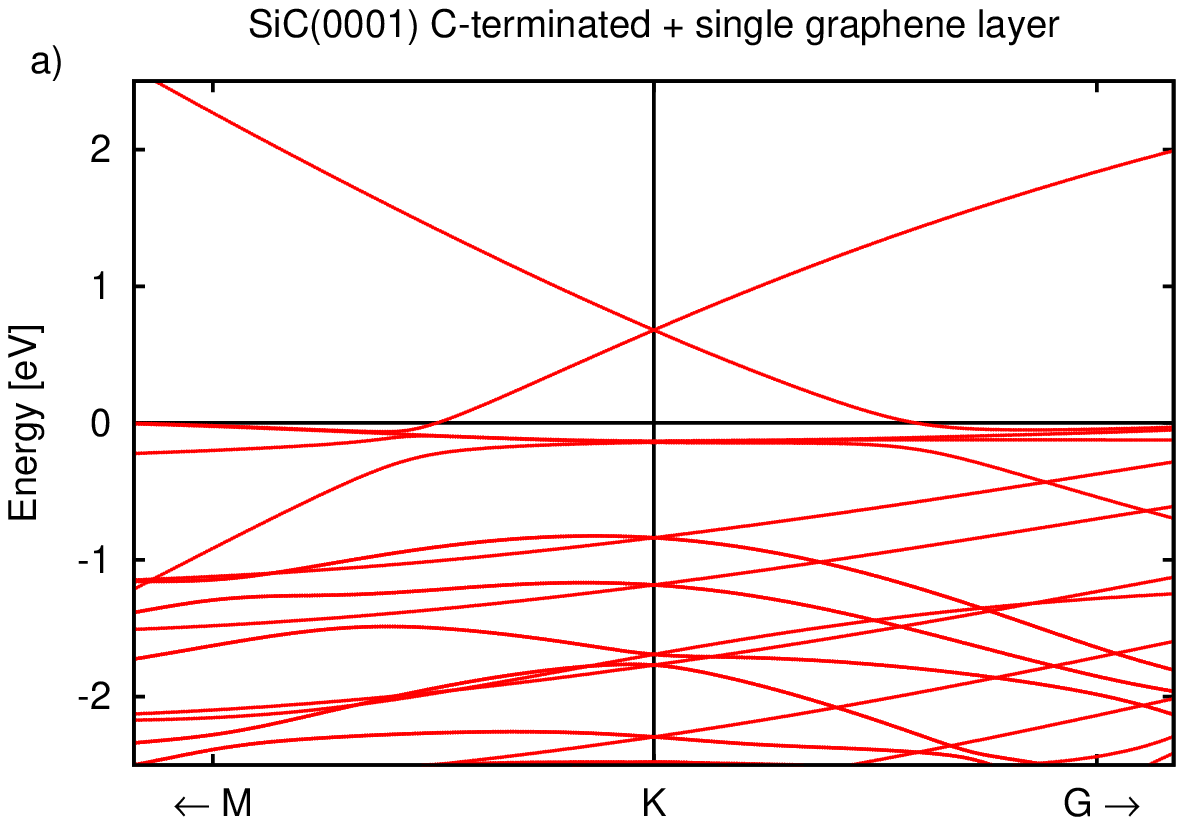} \hspace{2mm}
\includegraphics[scale=0.45]{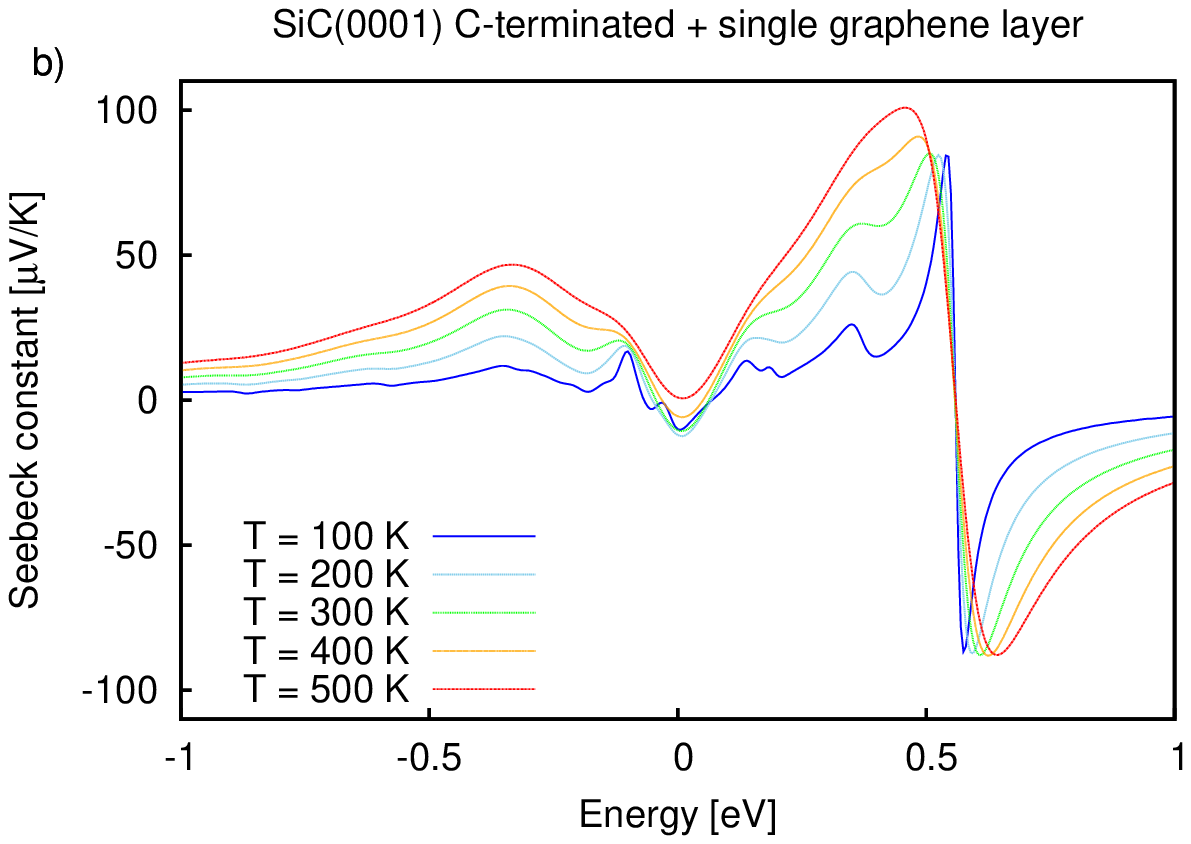}
\hspace{2mm} \includegraphics[scale=0.45]{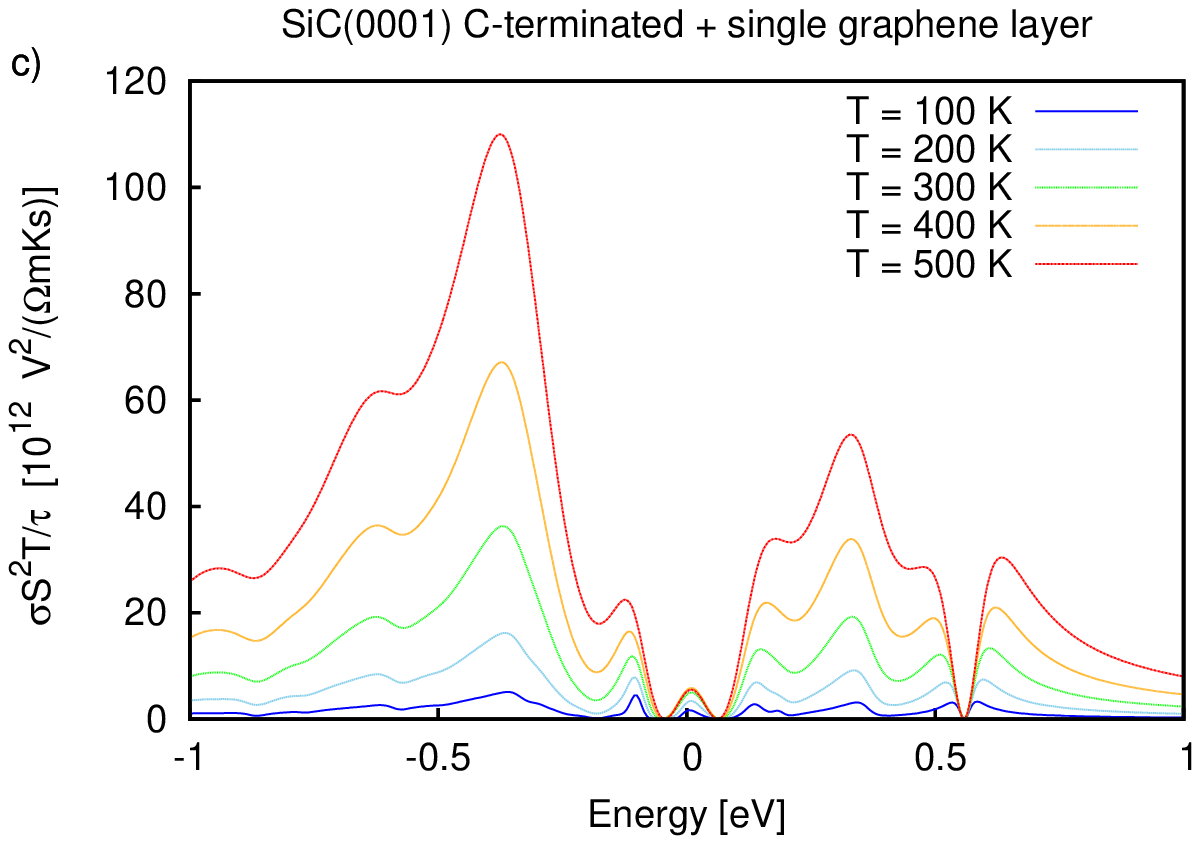} } \vspace{2mm}
 \leftline{\includegraphics[scale=0.45]{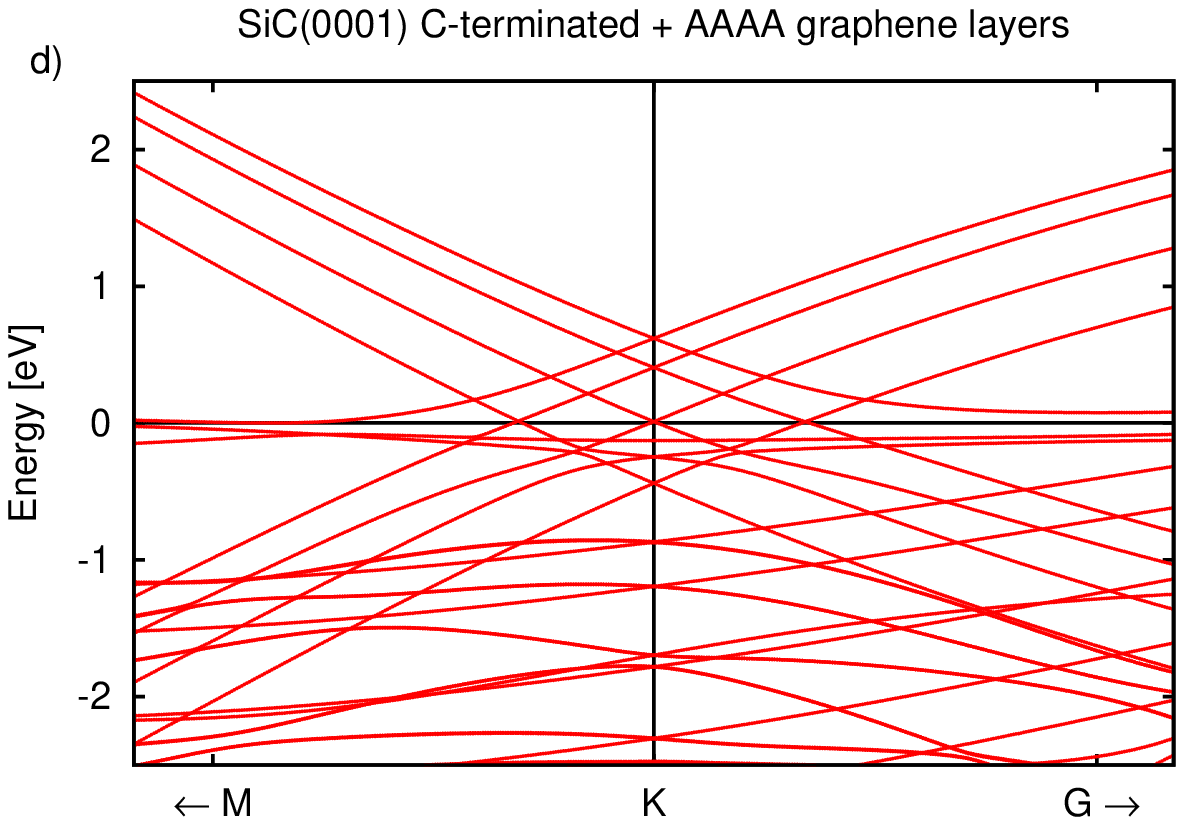} \hspace{2mm}
\includegraphics[scale=0.45]{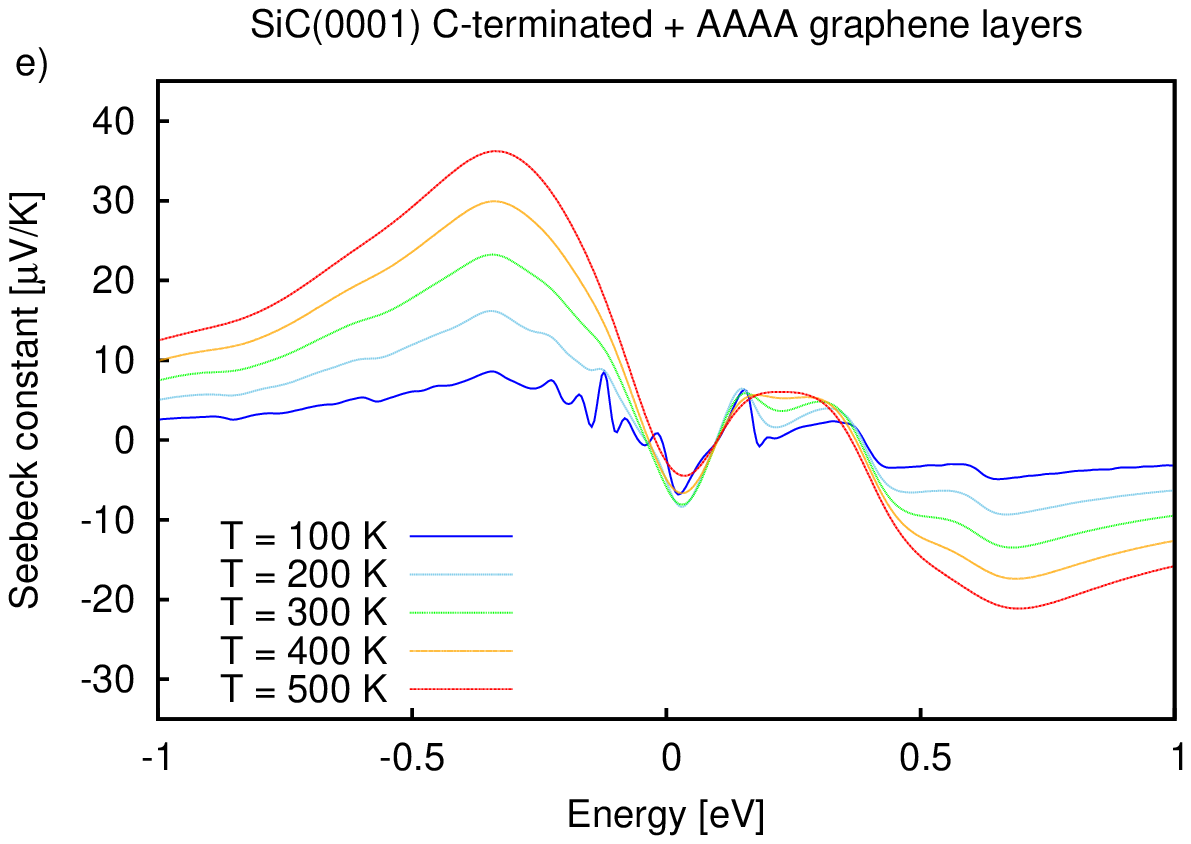}
\hspace{2mm} \includegraphics[scale=0.45]{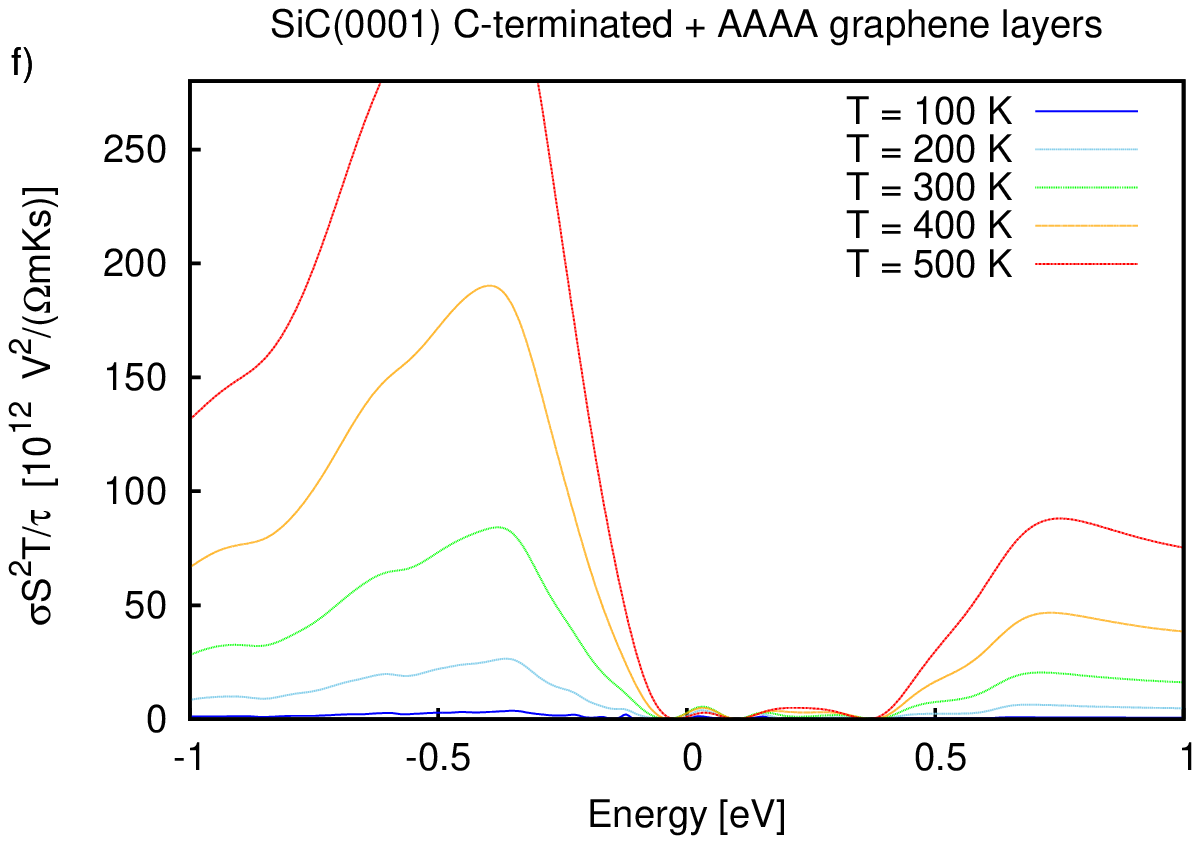} } \vspace{2mm}
\leftline{\includegraphics[scale=0.45]{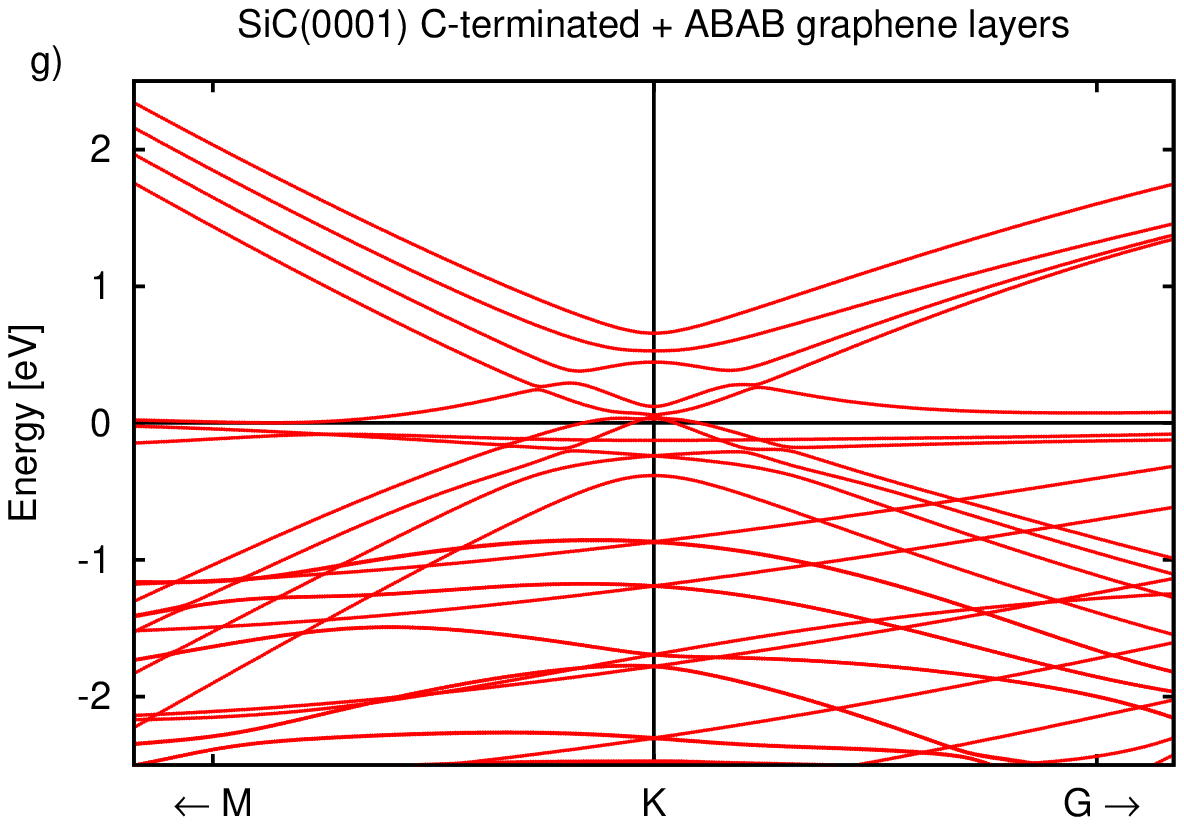} \hspace{2mm}
\includegraphics[scale=0.45]{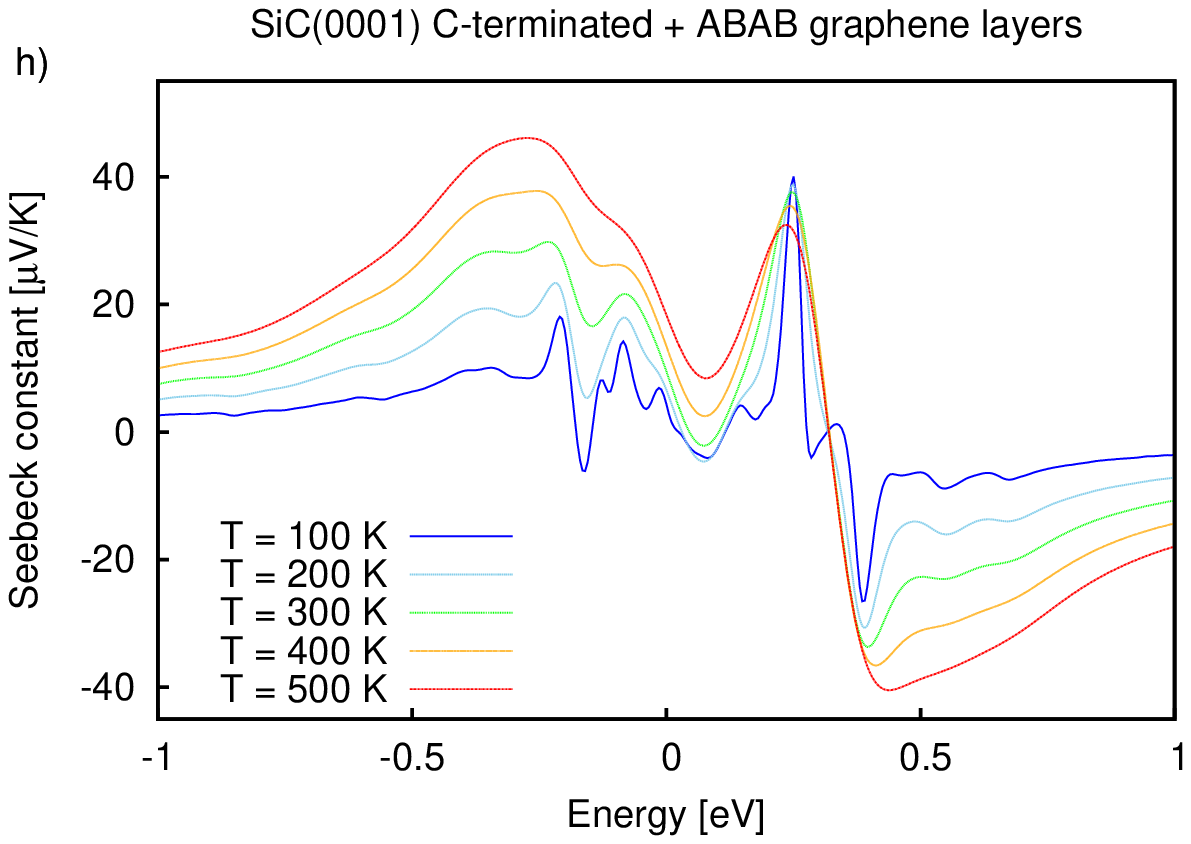}
\hspace{2mm} \includegraphics[scale=0.45]{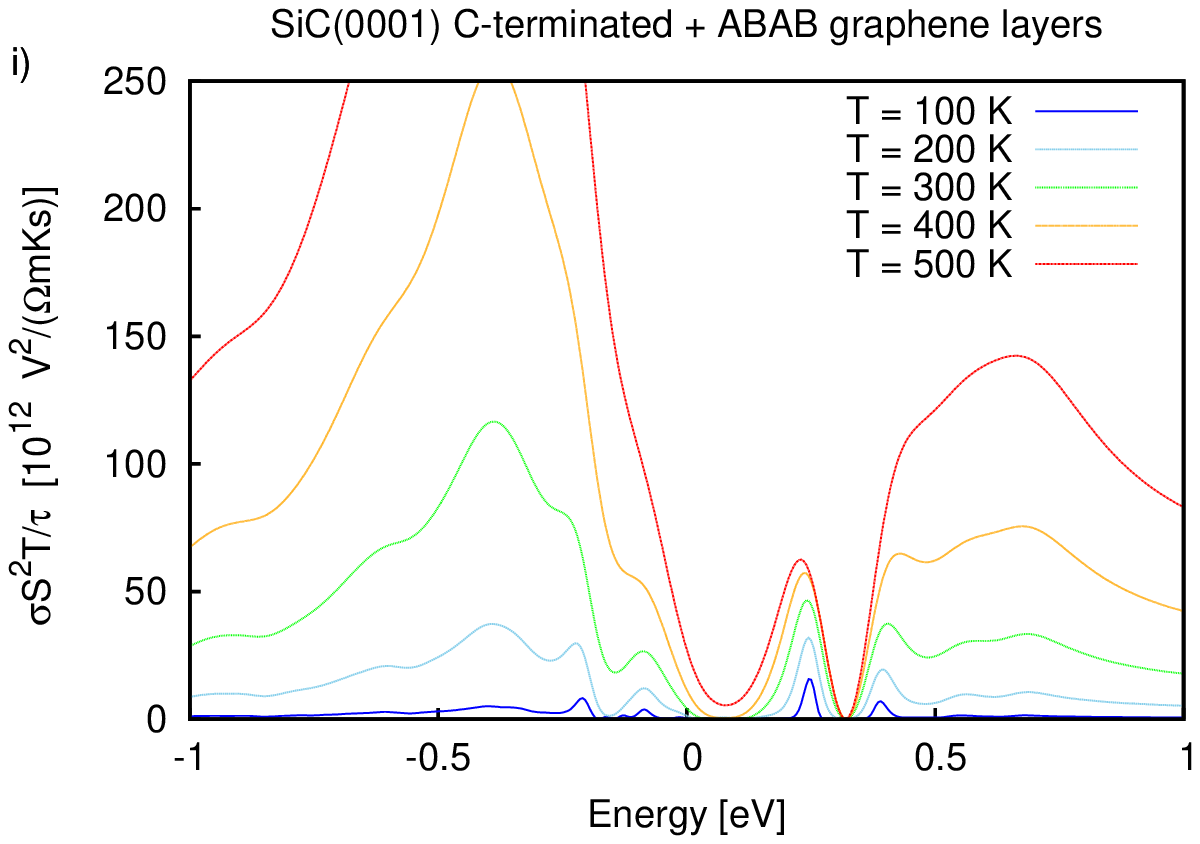} } \vspace{2mm}
 \leftline{\includegraphics[scale=0.45]{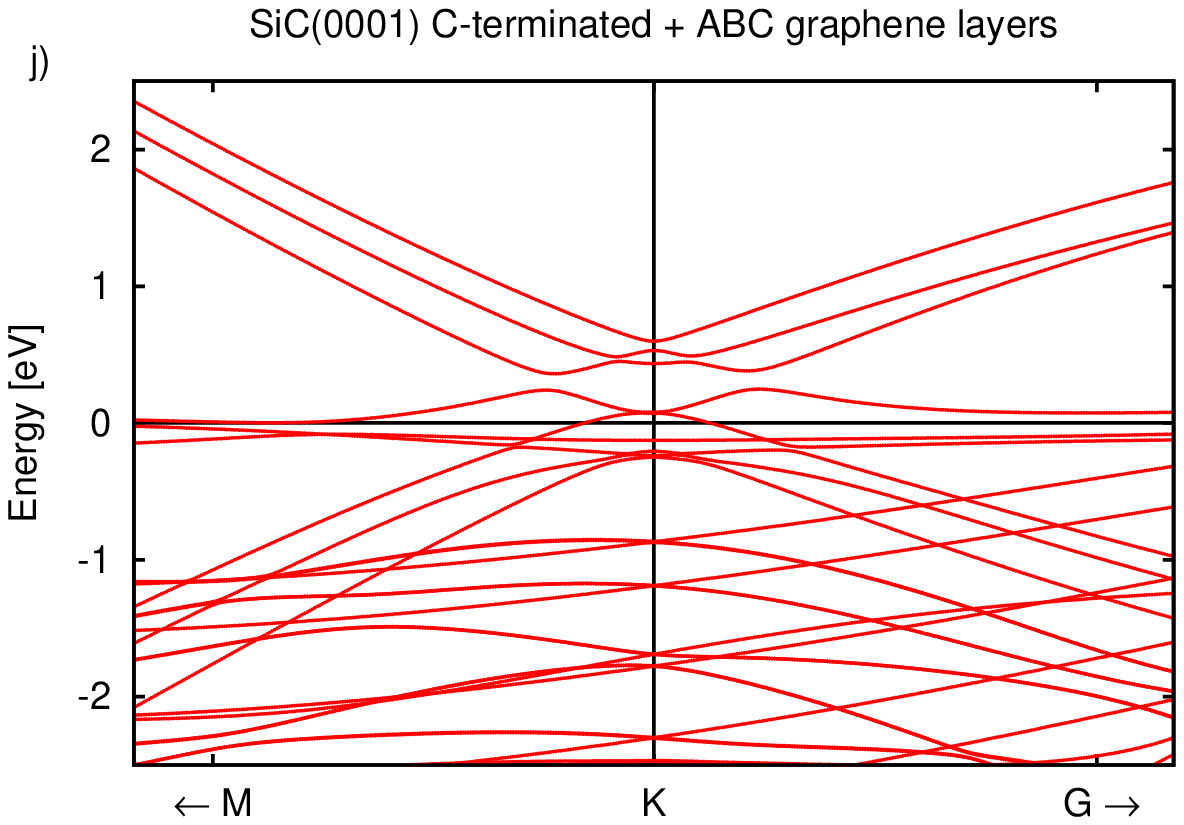} \hspace{2mm}
\includegraphics[scale=0.45]{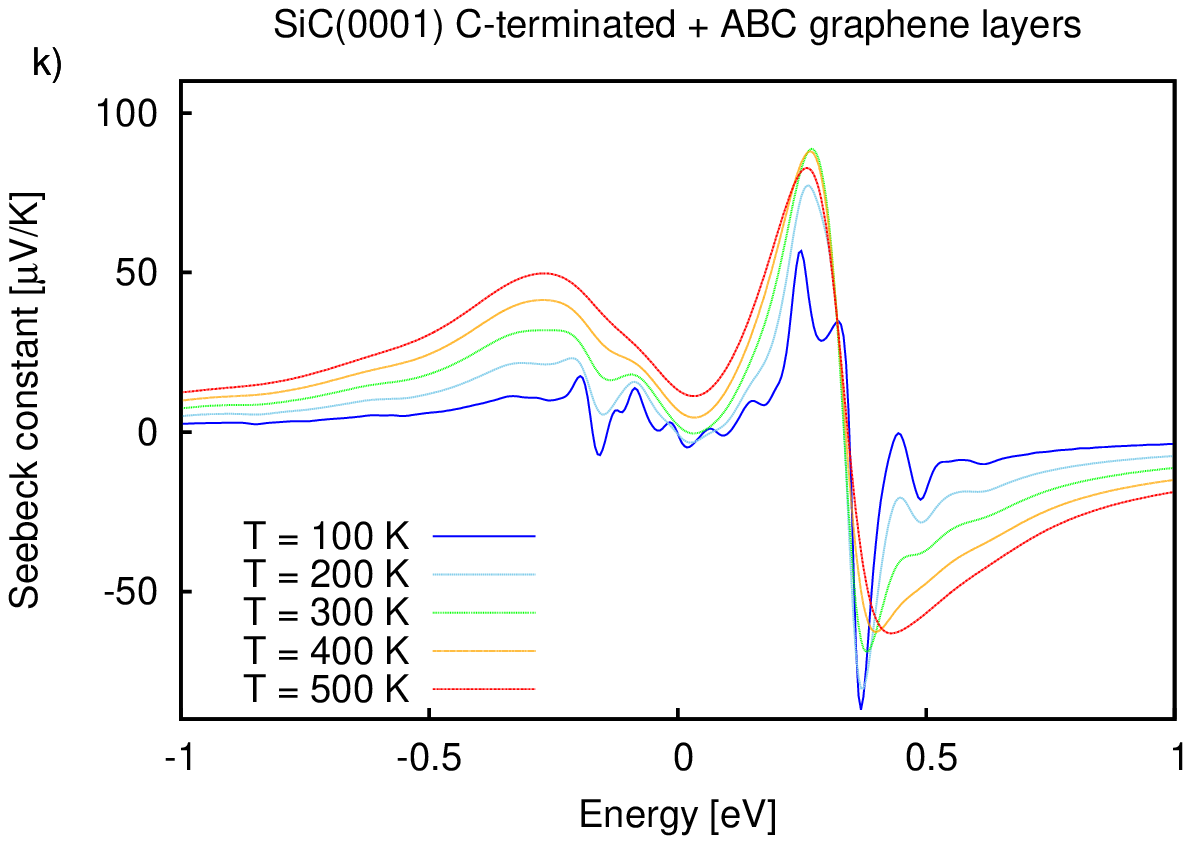}
\hspace{2mm} \includegraphics[scale=0.45]{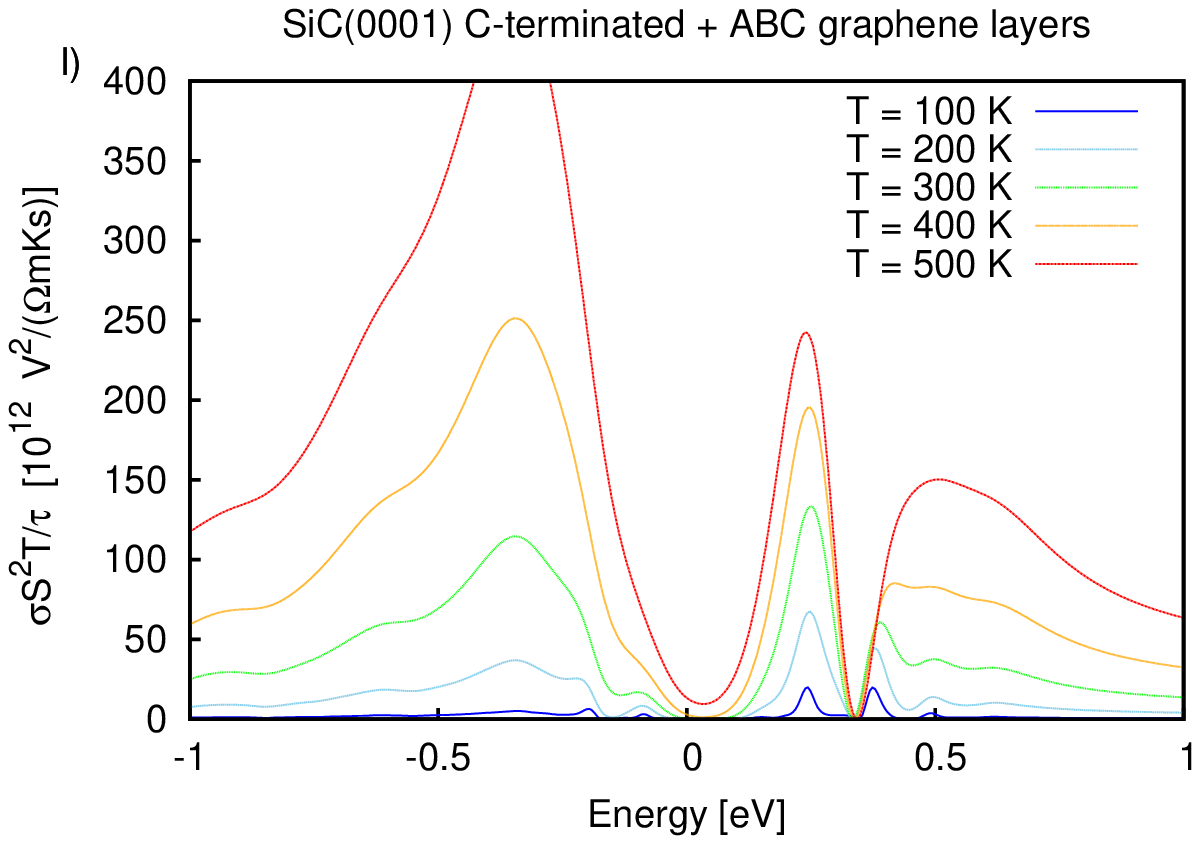} }
\caption{Band structure, Seebeck coefficient
and $\sigma S^2 T/\tau$ for graphene (a-c), 4ML
graphene in the AA stacking (d-f), 4ML graphene in the AB stacking
(g-i), and 3ML graphene in the ABC stacking (j-l). In all
cases, the layers are deposited on the C-face of a SiC(0001) substrate.}
\label{b3}
\end{figure*}
\begin{figure}
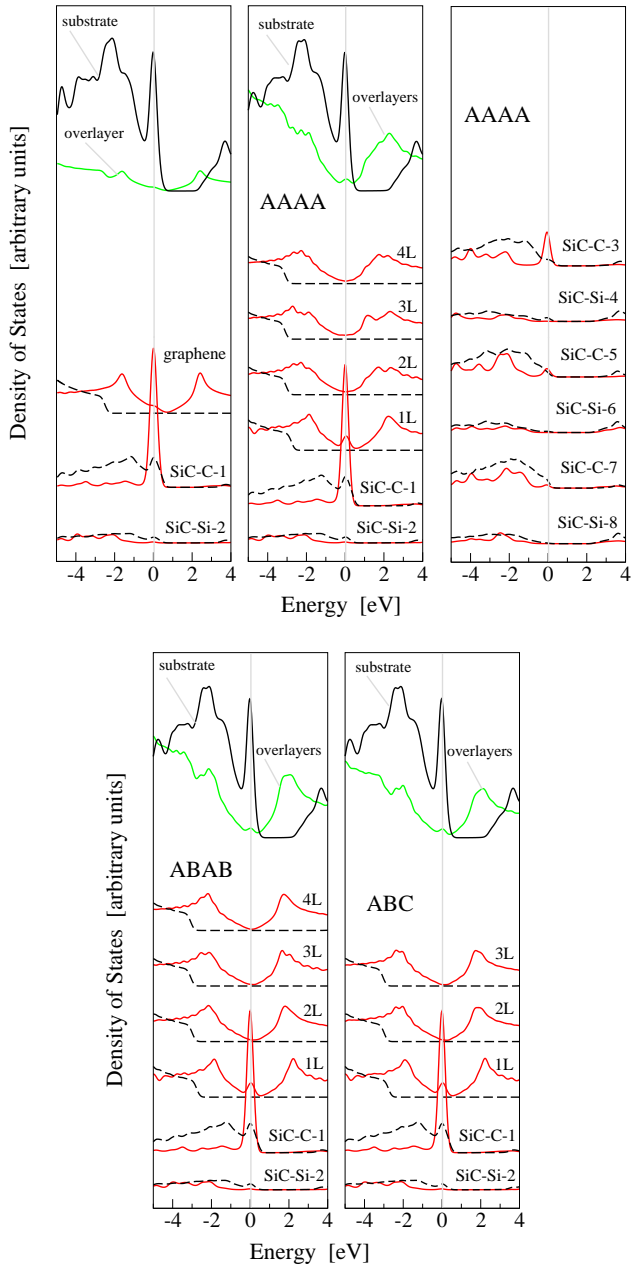

\leftline{\hspace{3.5cm}\includegraphics[scale=0.29]{F4-3}
\hspace{-8.5cm} \includegraphics[scale=0.29]{F4-1}} \vspace{4mm}
 \centerline{\includegraphics[scale=0.29]{F4-2} }
\caption{Projected local density of states for graphene
mono- and multilayers deposited on the C-face of SiC(0001). Red solid
lines denote the projection onto $p_{z}$ states, whereas black dashed
lines denote the projection onto $p_{x}$ and $p_{y}$ states. The
local DOS projected on each of the graphene layers,
as well as on the first two layers of the substrate, is reported.
For the case of the AA stacking, the projections onto
deeper substrate layers are also shown.
Total DOS of substrate and overlayers
is presented in the upper part of the panels.}
\label{b4}
\end{figure}

Since its discovery, graphene has been a top-list novel material thanks
to its remarkable properties.\cite{Barth,Castro} Beside extensive
experimental and theoretical basic studies, graphene has been recently
investigated for many novel device applications, like
2D-electronics and graphene-based transistors.\cite{transistors}
Nowadays, a limiting factor in the development of 
consumer electronics is the large chipset
heat generation; therefore, intense research\cite{Sanvito,Dresselhaus,Snyder} 
is being focused in minimizing the energy
losses due to heat generation or reusing the thermal energy by means of
thermoelectric devices.
The important thermal parameters
are the thermopower (or Seebeck coefficient) and the ZT figure of merit,
which must be as large as possible in order to maximize the efficiency of the 
thermoelectric conversion. Nanomaterials and composites 
are promising candidates since it is possible to engineer the 
phonon--phonon scattering, decreasing the thermal conductivity and
in turn increasing the ZT factor.\cite{ZT-nmat,ZT-arch,ZT-scirep}

In the case of graphene, after the growth process the layer can
be transferred onto an arbitrary substrate.\cite{transfer}
This substrate may be used as  a
source of electrical carriers (electrons or holes)
and may also transfer heat from the electrically functional
layer.\cite{Cu} In particular, due to the very good thermopower
properties of silicon carbide (SiC),
where the Seebeck coefficient exceeds $-480$~$\mu$V/K,
it has been shown that this substrate can be used as a
very good heat sink, performing in some cases even
better than a copper plate.\cite{Ssic,Cu}
Moreover, silicon carbide is especially convenient as a material for graphene
deposition, since it is a growth substrate for graphene monolayers
when the Si-terminated (0001) surface is used as substrate,\cite{Strup,Borys1}
or for graphene multilayers when the C-terminated face is used.\cite{Borys2}
Additionally, SiC doped with boron is a superconductor\cite{Bsic}
and can be used for graphene-superconductor junctions, where the Seebeck
coefficient can be strongly enhanced at specific temperatures in the Andreev
reflection regime.\cite{Spalek} Moreover, doping SiC with boron might
not only cause superconductivity but also increase
the thermoelectric efficiency as happens in many other
boron-containing materials.\cite{Boron}

The experimental and theoretical studies of the graphene structure on
SiC reveal interesting geometric phases: i) graphene monolayers grown
onto the Si-face of SiC have a strong buckling,\cite{Si1,Si2}
i.e. variable graphene to Si interatomic distances, ii) graphene grown
onto the C-face of SiC form flat multilayers which occur in different
stackings, i.e. orders of atoms in subsequent layers. Multilayers
can be classified in one of three families:\cite{Borys2} C atoms exactly on
top of each other (AA stacking), the Bernal stacking (AB stacking) and the
rhombohedral stacking (ABC stacking), as illustrated in Fig.~\ref{b1}.
Graphene layers interact among each other mainly via van der Waals interactions,
and different stackings exhibit distinct properties. 
For instance, the stacking order and the presence of defects
in graphene can be visualized via a measure of the thermoelectric
power.\cite{order}

In the literature, extensive studies have been focused on
evaluating and measuring the thermopower and the thermoelectric figure
of merit for pure graphene mono- and bilayers, often at the Si/SiO$_{2}$ substrate;
a comprehensive review collects these data.\cite{grev} However detailed
knowledge of these parameters for various stackings and number of
multilayers is missing, and the effect of the SiC substrate also has not
been investigated yet, except in the case of a single graphene layer.\cite{expCSiC}

Therefore, we report here theoretical calculations of the Seebeck coefficient and
of the electrical contribution to the ZT figure of merit 
for different mono- and multilayers. 
We examine both the free-standing case, 
and the case of layers deposited on top of a SiC substrate, 
which we assume to be the pure, undoped, semi-insulating crystal. 
The graphene layers are modelled as if they were directly grown on the substrate. Therefore,
no damage of the periodicity or intercalation with water or any gas was taken into account.   
We performed the calculations for various temperatures and for 
three different possible stackings: AA, AB and ABC. 
Other longer-range orders can exist, such as e.g. the turbostratic order,
characterized by a rotation of subsequent layers by a certain angle.\cite{turbo}
Such long-range periodicity needs to be modelled with large cells 
and is beyond the scope of this work; however, we can expect that
band structures (and, accordingly, the thermoelectric properties)
would be a superposition of the investigated short-range orders. \\

Calculations were performed within the density-functional theory (DFT)
framework with the PW91 approximation for the exchange--correlation
functional,\cite{pw91} 
as implemented in the \textsc{Quantum ESPRESSO} suite of codes\cite{qe}
which adopts a plane-wave basis set and uses pseudopotentials to approximate
the core region. We used the PW91 ultrasoft pseudopotentials freely
available from the \textsc{Quantum ESPRESSO} website, with energy
cutoff of 20 Ry for the wavefunctions and 200 Ry for the charge density.
The experimental geometry was used for the interlayer distances:
3.2~\AA{} for the first graphene layer above the SiC substrate,
3.7~\AA{} between the first and the second graphene layer, and 3.4~\AA{} 
for each subsequent layer.\cite{Borys2} The SiC substrate
(when present) was simulated using 10 monolayers, more than sufficient
to properly approximate a semi-infinite bulk, with an in-plane atomic
arrangement for the first graphene monolayer on the C face of the
SiC surface that fits 
a $\sqrt{3}\times\sqrt{3}R30^{\circ}$ supercell with respect to the
SiC surface atoms.\cite{footnote}
The bottom-face of the substrate has been passivated with H atoms.
The Brillouin zone has been sampled using a $20\times20\times1$ $\Gamma-$centered
Monkhorst-Pack uniform k-mesh. The vacuum separation between periodic
slabs in the direction perpendicular to the surfaces has been kept to about
50~\AA{} in order to avoid interactions between periodic images.
After the DFT calculations, we used the Wannier90 code\cite{w90-2}
in order to obtain the maximally-localized Wannier functions
(MLWFs)\cite{RMP}.
We minimized the spread of the MLWFs using atom-centered orbitals as
starting guess:
the $p_z$ orbital on each C atom and
hybridized $sp^2$ orbitals on every other C atom
 of the graphene layer;
and $sp^3$ orbitals on every Si atom of the SiC substrate.
The frozen energy window was chosen from the bottom of the valence
band to the Fermi
energy, and the outer energy window 30~eV above the Fermi level.
We verified that the obtained MLWFs were real-valued.
We then used the MLWFs to interpolate the band structure on a very dense
$k-$mesh ($2400\times2400$ points in the graphene plane for free-standing cases
and $1600\times1600$ points for the deposited cases, which are characterized by 
smaller Brillouin zones). Such a dense
mesh is necessary for an accurate description of the band structure
of the systems under investigation, as we also discuss later, and
is possible thanks to the extremely accurate and fast interpolation of the
electronic bands in a maximally-localized Wannier functions basis set.\cite{NM}
Finally, we calculated the transport distribution function and the
other transport properties presented here using the semiclassical
Boltzmann equations in the constant relaxation time approximation
using the \textsc{BoltzWann} post-processing code~\cite{BW}
distributed with Wannier90 v.2.0.\\

The band structures for bulk graphite, free-standing monolayer graphene
and multilayer graphene in the AA, AB and ABC stackings are compared
in Fig.~\ref{b1}, where we also show the atomic arrangement and the
calculated Seebeck coefficients $S$ for different temperatures and number
of monolayers (ML). We emphasize here that, in the constant relaxation
time approximation, $S$ is independent of the value of the
relaxation time $\tau$.
For all the free-standing systems, the Seebeck
coefficient as a function of the chemical potential displays the electron-hole
symmetry, at least in a range of about 1.5 eV around the Fermi energy.
This symmetry is also revealed in experiments for deposited graphene
with small doping rates, i.e. close to the Fermi level of the undoped
system\cite{symm}, and is particularly useful since it allows to
determine whether a graphene layer is $p-$ or $n-$type doped by means
of a thermopower measurement.\\

Our results show that the maximum of the Seebeck coefficient at T=300
K achieves 86 $\mu$V/K for graphene, 66 $\mu$V/K for graphite, 12
and 69 $\mu$V/K for 4-monolayers (4ML) in the AA- and AB-stackings
respectively, and 71 $\mu$V/K for three monolayers (3ML) in the ABC-stacking.
Inspecting the graphene band structure near the Dirac point, we notice
that ``V-shaped'' bands give a negative contribution to the Seebeck
coefficient, while the ``A-shaped'' bands give a positive contribution.
This explains why the AA stacking shows suppressed Seebeck effect with
respect to other types of stackings. The tiny details of complex ``V-A-shaped''
band structure in the ABC stacking are also reflected in the oscillating
Seebeck coefficient at low temperatures. We emphasize that we verified
by increasing the $k-$mesh that these oscillations are a real physical
effect and not the result of numerical noise. We also note that a
change in the number of monolayers (ML) in free-standing graphene significantly
influences the thermoelectric response only for the AA stacking.
This is due to the stronger interactions between
carbon atoms stacked on top of each other with respect to 
interactions between monolayers in the
AB and ABC stackings. On the other hand, the AB stacked graphene is
thermoelectrically similar to the single graphene monolayer and, as
expected due to the similar stacking, to graphite.

The values that we obtain are in good agreement with experiments for
samples mechanically exfoliated on Si/SiO$_{2}$. For instance, a thermopower
(TEP) value of about 93 $\mu$V/K is reported in 
Zuev et al.~\cite{expS} for
p-type graphene at 300 K under a gate voltage of $V_{g}=-5$V. A maximum
TEP value of around 95 $\mu$V/K at 280 K is reported in 
Wei et al.~\cite{expS3}
Other authors obtain 40 $\mu$V/K close to V$_{g}$=0V at 255 K, with
a maximum around 50 $\mu$V/K near V$_{g}$=25V\cite{expS2}. Our
results also agree with previous model calculations for graphene,
that report a TEP close to 80 $\mu$V/K at 300~K.\cite{Sarma}

The efficiency of a thermoelectric material, however, does not depend
uniquely on the Seebeck coefficient, but is instead described by the
thermoelectric figure of merit ZT, defined as $ZT=\sigma S^{2}T/\kappa$,
where S is the Seebeck coefficient, $\sigma$ is the electrical conductivity,
and $\kappa=\kappa_{e}+\kappa_{l}$ is the thermal conductivity, 
having an electronic contribution $\kappa_e$ and a lattice contribution $\kappa_l$. 
In many systems of interest at room temperature, $\kappa_l$ is the leading term. 
The evaluation of $\kappa_l$ requires however the evaluation of phonon-phonon 
scattering terms, that is beyond the scope of this work. 
We therefore show in the following figures the quantity $\sigma S^2 T/\tau$, 
that is independent of the value chosen for relaxation time $\tau$, 
and is an indicator of the behavior of the electronic contribution 
to the $ZT$ coefficient. \\

The calculated values for free-standing structures in all stackings are
shown in Fig.~\ref{b2} in the $200-400$~K temperature range and
for different number of monolayers. The maximal value of  $\sigma S^2 T/\tau$ 
increases monotonically with temperature for all stackings. We emphasize that this 
is not only due to the factor $T$, because the increase is more than linear 
(compare for instance the curves at 200~K and 400~K). 
Moreover, for the AB and ABC stackings, the dependence of $\sigma S^2 T/\tau$ 
partially loses the electron--hole symmetry. 
As already discussed above, interlayer interactions in the AA stacking
strongly suppress the Seebeck coefficient and, in turn,
the $\sigma S^2 T/\tau$
factor, especially for values of the chemical potential near the Fermi energy.
On the other hand, both the AB and ABC stackings show an increase 
of $\sigma S^2 T/\tau$ as the number of layers increases, with peaks located 
at both sides of the Fermi energy. This is particularly relevant for applications, 
showing that an increased thermoelectric efficiency could be achieved if multilayer 
systems (with the proper stacking) are considered, mainly thanks to the 
increased number of transport channels available and therefore to an increase of 
the electrical conductivity.
In particular, at 300 K, the maximum value reached by $\sigma S^2 T/\tau$ in the case of 
the AB stacking is 1.22, 2.67 and 4.07 times larger than the maximum of graphene 
for 2, 4 and 6ML, respectively. 
In the case of the ABC stacking, the corresponding
numbers are 2.18, 4.66 and 6.85, for 3, 6 and 9ML, respectively. \\

The effect of the deposition on the C-face of a SiC(0001) substrate
on the properties of graphene and its multilayers is technologically
extremely relevant. For this reason, we also investigate here the transport
properties of the same systems after deposition onto such a substrate.
In Fig.~\ref{b3}, the band structures, the Seebeck coefficient $S$
and the $\sigma S^2 T/\tau$ contribution to the thermoelectric efficiency
are reported for all investigated stackings at different temperatures.
The flat interface states are visible in the band structures and appear 
in all cases on the hole side.

In order to better understand the origin of the different bands
and features in the transport spectra, we also report in Fig.~\ref{b4}
the projected density of states (PDOS)
for all structures, where we
show the projection onto the $p-$like states of each graphene layer and of
the first two layers of the substrate.
In all cases, a fully occupied high density peak is present, originating
from the $p_{z}$ states of the topmost C-layer of the C-face SiC(0001)
substrate, corresponding to the almost flat bands just below $E=0$.
A broader maximum, also of $p_{z}-$type but on the
unoccupied side, arises from the graphene layers.
In order to understand whether the most relevant contributions to the Seebeck coeffiecient
and to $\sigma S^2 T/\tau$ originate mainly from the overlayers or from the substrate, we
also show the total density of states decomposed in its contributions from
the substrate and the overlayers. \\

The shape of the total density of states (DOS) projected on  
the substrate is practically independent of the overlayer stacking, 
and moreover deep substrate layers have negligible contribution near the Fermi energy.
Therefore in Fig.~\ref{b4}, we show the projection on a few more
substrate layers only for the AAAA case. 
In particular, we can notice that
deeper C-layers of the substrate contribute much less to
the peak at the Fermi level. On the other hand, the Si layers show only
a small and broad contribution to the DOS. 
These results suggest that 
replacing the $4H$-SiC(0001) substrate by a $6H$- or even a $2H$-SiC(0001) 
would not change much the results on the thermoelectric properties 
that we discuss in this paper.

The L\"owdin analysis shows in all cases a similar charge distribution
over the substrate and graphene multilayers: the substrate Si atoms
have a net charge of $-1.2$ electrons per atom with respect to their
valence charge of four; 
the substrate C atoms have a net charge of $+1.1$, except for
the topmost C layer where each C atom has a charge of +0.8. 
The C atoms in the graphene multilayers have a net charge in the range of
$[-0.06,-0.04]$, with the smallest net charge (in absolute value)
reached for the central layers of the multilayer graphene.
We emphasize here that the charge disproportion effect that
we find is not very pronounced, at least for the thin multilayers
(a few ML) that we investigated.
In Lin et al.,~\cite{3trans} instead, three different types of electrical
conductivities --- electron, intrinsic, and hole ---
have been reported for multilayer graphene sheets,
indicating that the position of the Fermi level is layer-dependent.
However, in their experiments, the number of
graphene MLs is expected to be larger than 10.
It is therefore possible that, when increasing
the number of ML, the charge disproportion effect becomes more important,
giving rise to different spatial regions that are effectively neutral,
$n-$ or $p-$doped.

In Fig.~\ref{b3}, the Seebeck coefficients show two main oscillatory behaviors,
one occurring near the Fermi energy where the localized states are present,
and one at the energy at which the Dirac cone is localized. Indeed, as already
discussed using the results of Fig.~\ref{b4}, the PDOS for energies above 
the Fermi level is almost completely due to the graphene overlayers, 
and indeed the Seebeck coefficient for positive $\mu$ resembles the corresponding 
spectrum of the free-standing case (after having taken into account a shift of 
the Fermi energy upon deposition).
However, if we focus on the $\sigma S^2 T/\tau$ contribution,
our results show that a very strong enhancement is obtained for ABC stacking
upon deposition on SiC.
In fact, at 300 K, the maximum value (for $\mu > 0$) 
reached by $\sigma S^2 T/\tau$ in the case of
deposited graphene is 0.95 of  the maximum of the same layer
in the free-standing case. The same ratio of the deposited with respect to 
the free standing AAAA (4ML) is 0.37,  for ABAB (4ML) it is 0.97,
and for ABC (3ML) it is 2.99. 
Interestingly, in the latter case (ABC stacking)
we obtained a band gap opening just above the Fermi level
as a result of the interactions between the overlayers 
and the substrate. The thermoelectric parameters of the material 
are enhanced as a consequence of the
increased number of band edges and thus the vanishing
band velocities.\cite{MWAD}

The thermopower coefficient calculated in this work for graphene deposited 
on SiC achieves a maximum value of 86 $\mu$V/K for a temperature of 230 K,
to be compared with the experimental value reported
in Wu~\cite{expCSiC} for a single hole-doped 
graphene layer on the C-face
of the same substrate, and at the same temperature, which is
55 $\mu$V/K. The difference is large but within our theoretical range.
For a comparison
with the deposition on different substrates, the measured TEP for
graphene deposited on SiO$_{2}$ substrate is about 20~$\mu$V/K
for $p-$type or $-50$~$\mu$V/K for $n-$type samples.\cite{SiO2}
For multilayered
graphene deposited on Si/SiO$_{2}$, there are several experimental
reports showing values ranging from 40 $\mu$V/K up to even 700 $\mu$V/K
for gapped graphene functionalized with molecules.\cite{M1,M2}
The upward shift of the Dirac cone in graphene deposited on a substrate is an
usual phenomenon, which is also visible in thermoelectric experiments for 
graphene grown at SiO$_2$ and published by Novoselov et al.\cite{Novoselov-Sci}
(see Fig.~2 in that paper). Another important issue to take into account is the fact that
we assumed a perfect in-plane periodic structure in the calculations,
but in the experiments the buffer layers between the substrate and
the perfect graphene layer are discontinuous, island-like shaped.\cite{Kacper}
In this case, local interface states would
show the properties of the quantum-dot states (i.e., flat bands).
On the other hand, it is difficult to predict the
step-end effects and their interactions with the 
overlayers and the substrate. Such edges contain the unpaired electrons, so that
donor states would appear above the Fermi level, and there would be 
and increased chemical reactivity with the atmosphere gases and water.

Summarizing, we calculated the DFT band structures of graphene mono-
and multilayers both free-standing and deposited on the C face of
 $4H$-SiC(0001).
The corresponding thermoelectric properties (Seebeck coefficient and 
$\sigma S^2 T / \tau$) were evaluated
adopting the semiclassical Boltzmann equations in the constant relaxation
time approximation, where the electronic bands were interpolated using 
a maximally-localized
Wannier functions basis set. The Seebeck coefficient
is similar for the AB and ABC stackings and monolayer graphene
in the free-standing case. The AA stacking is instead thermoelectrically
much less efficient. The effect of deposition on SiC strongly increases
the $\sigma S^2 T/\tau$ contribution to ZT in the case of the ABC stacking.
For the AA stacking, instead, the interactions
between the graphene layers and with the C-face of the SiC substrate reduce
the parameters of interest. 
These results are optimistic for a design
of graphene and SiC-based heterostructures as future electronic devices
and provide an indication of which type of stacking can be the most
suitable.

This work has been supported by the European Funds for Regional Development
within the SICMAT Project (Contract No. UDA-POIG.01.03.01-14-155/09).
Calculations have been performed in the Interdisciplinary Centre of
Mathematical and Computer Modeling (ICM) of the University of Warsaw,
within the grants G51-2 and G47-7, and partially supported by PL-Grid
Infrastructure.


\begin{thebibliography}{10}
\bibitem{footnote}
We did not perform calculations of graphene
deposited on the Si-face of the SiC substrate because they would require
a much larger supercell.
\end{thebibliography}
\end{document}